\documentclass[11pt,oneside,english]{amsart}

\usepackage[T1]{fontenc}
\usepackage[latin9]{inputenc}
\usepackage[letterpaper]{geometry}
\usepackage[active]{srcltx}
\usepackage{prettyref}
\usepackage{amsthm}

\makeatletter
\numberwithin{equation}{section}
 \theoremstyle{definition}
 \newtheorem*{defn*}{\protect\definitionname}
\theoremstyle{plain}
\newtheorem{thm}{\protect\theoremname}
  \theoremstyle{remark}
  \newtheorem*{rem*}{\protect\remarkname}
  \theoremstyle{plain}
  \newtheorem{lem}[thm]{\protect\lemmaname}
  \theoremstyle{plain}
  \newtheorem{cor}[thm]{\protect\corollaryname}

\usepackage{multicol} 
\usepackage{verbatim}

\makeatother

\usepackage{babel}
  \providecommand{\corollaryname}{Corollary}
  \providecommand{\definitionname}{Definition}
  \providecommand{\lemmaname}{Lemma}
  \providecommand{\remarkname}{Remark}
\providecommand{\theoremname}{Theorem}

\begin{document}

\title{A Threshold For Clusters in Real-World Random Networks}

\maketitle
\begin{center}
\textsc{Arron Norwell} \\
University of Toronto \\
anorwell@cs.toronto.edu \\
\end{center}
\thispagestyle{empty}
\begin{abstract}
Recent empirical work \cite{Leskovec2009} has suggested the existence
of a size threshold for the existence of clusters within many real-world
networks. We give the first proof that this clustering size threshold
exists within a real-world random network model, and determine the
asymtotic value at which it occurs. 

More precisely, we choose the \emph{Community Guided Attachment} (\emph{CGA})
random network model of Leskovek, Kleinberg, and Faloutsos \cite{Leskovec2005}.
The model is non-uniform and contains self-similar communities, and
has been shown to have many properties of real-world networks. To
capture the notion of clustering, we follow Mishra et. al. \cite{Mishra2007},
who defined a type of clustering for real-world networks: an $\left(\alpha,\beta\right)$\emph{-cluster}
is a set that is both \emph{internally dense} (to the extent given
by the parameter $\beta$), and \emph{externally sparse }(to the extent
given by the parameter $\alpha$) . With this definition of clustering,
we show the existence of a size threshold of $\left(\ln n\right)^{\frac{1}{2}}$
for the existence of clusters in the CGA model. For all $\epsilon>0$,
a.a.s. clusters larger than $\left(\ln n\right)^{\frac{1}{2}-\epsilon}$
exist, whereas a.a.s. clusters larger than $\left(\ln n\right)^{\frac{1}{2}+\epsilon}$
do not exist. Moreover, we show a size bound on the existence of small,
constant-size clusters.
\end{abstract}
\newpage{}

\setcounter{page}{1}

\section{Introduction\label{sec:Introduction}}

Real-world networks are everywhere. Examples include the network formed
by the connections between people in a city; the network of citations
between academic papers; an electric power grid; and the network of
physical interactions between proteins \cite{Newman2003}. Despite
their differing origins, emperical observation has shown these networks
to share many properties. These include: the \emph{small-world effect
}(the average shortest path between two nodes in a real-world network
is smaller than one might expect, and may shrink over time) \cite{Watts1998,Kleinberg2000};
the \emph{scale-free }property\emph{ }(the degree distribution of
nodes in the network follows a power law) \cite{Barabasi1999,Faloutsos1999};
and \emph{clustering }(certain parts of the network are much more
closely connected than their surrounding neighbourhood) \cite{Watts1998,Barrat2000}.
The present work concerns this last property of clustering. Clustering
is greatly important in biology, sociology, and computer science (see
e.g. \cite{Adamcsek2006,Abello2002,Kernighan1970,Nepusz2008,Newman2004,Newman2004a,Palla2005,Pinney2006};
\cite{Fortunato2010} lists hundreds of others), and plays an important
role in understanding the structure of real-world networks \cite{Watts1998,Adams2007,Barrat2000,Fortunato2010,Girvan2002}.
Despite this, our work is the first we know of to study clustering
analytically in any random model for real-world networks.

Clusters in real-world networks often overlap \cite{Mishra2007,Adamcsek2006};
that is, a single node may be a member of more than one cluster. One
can imagine that a computer in a computer network may belong to multiple
groups (corresponding to clusters); similarly, a person in a social
network may have multiple groups of friends. However, most approaches
to clustering partition the network without allowing for clusters
to overlap. In 2007, Mishra, Schreiber, Stanton, and Tarjan \cite{Mishra2007}
proposed the $\left(\alpha,\beta\right)$\emph{-cluster} as a new
formulation of clustering. In this definition, a cluster is a set
that is both \emph{externally sparse }(each node outside the set is
connected to only a few nodes outside the set, as determined by the
parameter $\alpha$) and \emph{internally dense} (each node in the
set is connected to many others inside the set, as determined by the
parameter $\beta$). This definition, motivated by real-world networks
such as social networks, allows for overlapping clusters.

A measure related to clustering is that of \emph{conductance. }The
conductance of a set is the ratio between the number of ``cut edges''
(the edges between the set and its complement) and the number of internal
edges in that set. A set with low conductance has many internal edges
and few ``cut'' edges, and is therefore intuitively a good cluster.
Recently, Leskovek, Lang, Dasgupta, and Mahoney \cite{Leskovec2009}
empirically examined how the conductance of the best conductance clusters
changed as cluster size increased. They found similar behaviour in
many existing real-world networks: below a certain size threshold,
good clusters exist; moreover, increasing cluster size below the threshold
improves the quality of the best cluster. Above the threshold, however,
increasing the cluster size decreases the quality of the best cluster. 

In other words, there appears to exist a size threshold for \emph{conductance}
clusters in real-world networks. In this work, we give the first proof
of a clustering size threshold in any model. More precisely, we choose
to work with $\left(\alpha,\beta\right)$-clusters. Because $\left(\alpha,\beta\right)$-clusters
were created specifically for types of real-world networks, they are
a natural formulation to choose. Instead of an empirical approach,
our method will be to study clustering analytically within an existing
model of a real-world random network. The benefits of this approach
are two-fold: first, it allows us to consider the asymptotic behaviour
of clusters as the network size grows; and second, it allows us to
prove directly the existence of a threshold.

Many random graph models that contain properties of real-world networks
have been proposed (e.g. \cite{Watts1998,Barthelemy1999,Aiello2000,Kleinberg2000,Bollobas2001b,Lattanzi2009}).
In 2005, Leskovek et. al. \cite{Leskovec2005} observed that real-world
networks obey the additional property of \emph{densification }(the
average degree of a node will grow over time), and proposed a model
called \emph{Community Generated Attachment} (CGA). Because this model
is built from a self-similar structure of nested communities, it exhibits
non-trivial clustering in a way that previous models do not. It was
also the first model to exhibit densification in addition to being
scale-free. Furthermore, its simple mathematical description makes
it amenable to analysis; other models that exhibit densification have
not tended to permit analysis \cite{Lattanzi2009}. For these reasons,
we choose to work with the CGA model, which will be defined fully
in \prettyref{sec:Model-and-Definitions}.

The goal of the present work is to analyze conditions under which
$\left(\alpha,\beta\right)$-clusters occur in the CGA model for real-world
networks. For every fixed $0<\alpha,\beta\leq1$, we establish a cluster
size threshold of $\left(\ln n\right)^{\frac{1}{2}}$ (where $n$
is the size of the network): for all $\epsilon>0$, a.a.s.%
\footnote{We say an event occurs \emph{asymptotically almost surely} (\emph{a.a.s.})
if the probability the event occurs approaches one as $n\rightarrow\infty$. %
} there are clusters larger than $\left(\ln n\right)^{\frac{1}{2}-\epsilon}$,
while a.a.s. there are no clusters larger than $\left(\ln n\right)^{\frac{1}{2}+\epsilon}$.
Furthermore, we show a size bound on the existence of small, constant-size
clusters.

Our work is is the first instance we know of that studies the existence
of any notion of clustering analytically in a random model for real-world
networks.

\section{Model and Definitions\label{sec:Model-and-Definitions}}

\subsection{CGA Random Graph Model}

The community-guided attachment (CGA) random graph model was first
proposed by Leskovek, Kleinberg, and Faloutsos \cite{Leskovec2005},
in response to the observation that real-world random graphs tend
to have average node degrees that increase over time, a property known
as densification. It was the first model proposed with this property.
Several other models with densification exist: Leskovek et. al. also
proposed the ``forest fire'' model \cite{Leskovec2005}; the model
in \cite{Leskovec2005a} can be shown to have densification and other
real-world network properties. Lattanzi and Sivakumar's recent affiliation
network model \cite{Lattanzi2009} is an example of a densifying model
that can admit analysis.

The CGA model is based on levels of nested, self-similar communities.
This is natural, because real-world networks often exhibit some level
of self-similarity. For example, a computer network may be decomposed
into several sub-networks, based on geography or purpose. Likewise,
each of these sub-networks may themselves be further decomposed into
smaller groups. A pair of computers sharing membership in one of these
small groups are much more likely to be connected than two computers
chosen at random from within the large network. This self-similar
hierarchical structure is also observable in other domains. For example,
it has been argued to apply to social groupings \cite{Watts2002},
subject classification of patents \cite{Leskovec2005}, and topic
classification of Web pages \cite{Menczer2002}. The CGA model itself
has previously been used to describe peer-to-peer networks \cite{Chen2009}.

The nested form of CGA makes it a natural choice of model to study
clustering. Random graph models using fixed power-law degree sequences
(e.g. \cite{MikeMolloy1997,Aiello2000}) are a.a.s. locally tree-like,
and hence exhibit no clustering at all. Methods such as Watts and
Strogatz' small-world model \cite{Watts1998} start with a regular
local graph structure such as a cycle or grid, and then add or re-wire
some number of edges at random to insert long-range edges. While these
models exhibit more clustering than a uniform random graph, it is
trivial clustering: clusters are determined by the original, deterministic
local structure rather than the random edges. In CGA, the nested communities
mean that small, more dense communities are more likely to contribute
to clusters than larger, less dense ones. As we will show, this implies
the existence of varied clusters of different sizes, formed by random
edges.

We now define the model precisely.
\begin{defn*}
Let $T$ be a complete tree of height $H$, with constant fan-out
$b$ (that is, each non-leaf node has exactly $b$ children). Let
$n=b^{H}$ be the number of leaves of $T$. We will construct a random
un-directed graph $G=\left(V,E\right)$ whose nodes $V$ are the leaves
of $T$. Given two nodes $u,v\in V$, we define the height $h(u,v)$
to be the height of the smallest subtree in $T$ that contains both
$u$ and $v$. (In other words, $h(u,v)$ is one half of the distance
between $u$ and $v$ in T.) For a parameter $c>1$, the probability
that our random graph $G$ will have an edge from $u$ to $v$ will
be equal to $c^{-h\left(u,v\right)}$.
\end{defn*}
The edge probability function chosen here is the only natural choice.
To see this, set $\Pr\left((u,v)\in E\right)=f\left(h\left(u,v\right)\right)$
for some function $f$. For $G$ to have a power-law degree sequence,
we require that $f\left(h\right)/f\left(h-1\right)$ is constant.
Hence, we must have $f\left(h\right)=\gamma c^{-h}$, where $c>1$
is the shrinking parameter, and $\gamma\leq1$ indicates the initial
density. For simplicity, we will take $\gamma=1$, but the work that
follows can be adapted for any value of $\gamma\leq1$.

Intuitively, each internal node of $T$ defines a sub-community of
$G$, given by the leaf nodes of the sub-tree of $T$ rooted at that
internal node. The larger the sub-tree, the larger and less connected
(on average) the sub-community.

We now define some terminology used within the CGA model:
\begin{defn*}
Let $M\subset V(G)$, so that $M$ corresponds to a subset of the
leaves of the tree $T$. Define the \emph{height} of $M$ to be the
height of the minimum complete subtree in $T$ containing all of $M$.
If a set $M$ has height $h$, then we will call $M$ \emph{complete}
if it has $b^{h}$ nodes. For each $h^{\prime}\geq h$, $M$ is a
subset of exactly one complete set of height $h^{\prime}$; we will
denote this set by $S\left(M,h^{\prime}\right).$ $S\left(M,h\right)$
is the \emph{minimum complete set containing $M$, }and will usually
be denoted $S\left(M\right)$. If $u$ is a leaf node of $T$ disjoint
from $S\left(M\right)$, notice that there exists a $j>h$ such that
for all $v\in S\left(M\right)$, $h\left(u,v\right)=j$. Therefore,
we will define $h\left(u,M\right)=j$ and refer to it as the \emph{height
of $u$ from $M$}.
\end{defn*}
Intuitively, the height of a pair of (leaf) nodes gives a prediction
of the similarity between those two nodes. Notice that the height
of a set is the maximum height over all pairs of vertices in the set. 

The above definitions are for the undirected version of the CGA model.
The directed CGA model is only a minor modification from this: instead
of a single edge between two vertices $u$ and $v$, we will have
two edges going in opposite directions, which occur indendently with
equal probability $c^{-h(u,v)}$. This version was the one originally
proposed by Leskovek et. al. \cite{Leskovec2005}. In this paper,
we work with the undirected version, but all conclusions hold in both
forms of the model with only trivial modifications.

\subsection{$\left(\alpha,\beta\right)$-clusters}

Intuitively, a cluster is a set of vertices that is more edge-dense
than the graph average. However, the exact definition chosen often
varies with application. For the real-world networks we consider,
allowing overlapping clusters is natural \cite{Mishra2007,Adamcsek2006}.
In other words, a single vertex should be allowed membership in more
than one cluster. Furthermore, it may be possible that a vertex is
not a member of any clusters at all. In 2007, Mishra et. al. \cite{Mishra2007}
proposed a clustering definition for social networks (and applying
to other real-world graph applications) that allows clusters to overlap.
This definition is as follows:
\begin{defn*}
For an undirected graph $G$, let $v\in V\left(G\right)$ and $M\subset V\left(G\right)$,
and let $e\left(v,M\right)$ denote the number of edges between $v$
and $M$. For parameters $0\leq\alpha,\beta\leq1$, \emph{we say $M$
is internally dense }if for all $v\in M$, $e\left(v,M\right)\geq\beta\left|M\right|$,
and is \emph{externally sparse }if for all $u\notin M$, $e\left(u,M\right)\leq\alpha\left|M\right|$.
An $\left(\alpha,\beta\right)$\emph{-cluster} is a set that is both
\emph{internally dense} and \emph{externally sparse}.
\end{defn*}
We will often refer to a $\left(\alpha,\beta\right)$-cluster as just
a cluster, with the parameters $\alpha$ and $\beta$ being implicit.
Note that for a cluster to necessarily be connected, we would have
to require $\beta\geq\frac{1}{2}$. For this reason, Mishra et. al.
restrict $\beta$ to this range, but nothing in this work places any
restriction on $\beta$. As well, it is natural to have $\alpha\leq\beta$,
but we do not require this either.

Within the CGA model, clusters may be described with the same terminology
as vertex sets:
\begin{defn*}
A cluster of height $h$ is a \emph{complete cluster} if it has $b^{h}$
vertices.
\end{defn*}
By using outgoing edges, the definition of $\left(\alpha,\beta\right)$-cluster
carries over to directed networks:
\begin{defn*}
For a directed graph $G$, let $v\in V\left(G\right)$ and $M\subset V\left(G\right)$,
and let $e_{d}\left(v,M\right)$ denote the number of edges from $v$
to $M$. For parameters $0\leq\alpha,\beta\leq1$, \emph{we say $M$
is internally dense }if for all $v\in M$, $e_{d}\left(v,M\right)\geq\beta\left|M\right|$,
and is \emph{externally sparse }if for all $u\notin M$, $e_{d}\left(u,M\right)\leq\alpha\left|M\right|$.
A \emph{directed }$\left(\alpha,\beta\right)$\emph{-cluster} is a
set that is both \emph{internally dense} and \emph{externally sparse}.
\end{defn*}
Note that the directed form of the clustering definition depends on
only outgoing edges. This corresponds to situations where cluster
membership depends only on outgoing intent (for example, online social
networks where users may ``subscribe'' to other users). While we
will work in the undirected CGA model using the undirected definition
of clustering, our results hold as well for the directed CGA model
using the directed definition of clustering.

\section{Current Work}

Our goal is to establish a size threshold of $\left(\ln n\right)^{\frac{1}{2}}$
for the existence of clusters in both the directed and undirected
versions of the CGA model. We will work in the undirected version
of the model, but all statements hold with trivial modifications in
the directed model. Our main theorem is stated as follows:
\begin{thm}
\label{thm:Thm1}Let $G$ be a graph chosen according to the undirected
CGA model. Then for all $0<\alpha,\beta\leq1$ and $\epsilon>0$:
\begin{description}
\item [{a}] Let $m^{\star}=\frac{\ln b}{\alpha\ln c}$. There a.a.s. exists
$\left(\alpha,\beta\right)$-clusters of size larger than $\left(\ln n\right)^{\frac{1}{2}-\epsilon}$
in $G$. Moreover, there exists a constant $\gamma=\gamma\left(\alpha,b,c\right)>0$
such that for each $h$ satisfying $\log_{b}m^{\star}<h\leq\left(\frac{1}{2}-\epsilon\right)\frac{\ln\ln n}{\ln b}$,
there a.a.s. exists at least $\left(\ln n\right)^{\gamma b^{h}}$
complete $\left(\alpha,\beta\right)$-clusters of size $b^{h}$.
\item [{b}] There are a.a.s. no $\left(\alpha,\beta\right)$-clusters with
more than $\left(\ln n\right)^{\frac{1}{2}+\epsilon}$ vertices.
\end{description}
\end{thm}
We will try to give some intuition towards why such a threshold might
occur. It turns out (see \prettyref{lem:E_1_and__E2-1}) that complete
sets of increasing (i.e., non-constant) size tend to be externally
sparse; we will attempt to intuitively justify why sets below the
threshold are internally dense.

To give intuition, we therefore make the following simplifications.
First, given a set $M$ of fixed size $b^{h}$, $M$ is more likely
to be internally dense if its height is small, because short-range
edges are more likely to occur. In the most extreme case, $M$ has
height $h$ and size $b^{h}$, and forms a complete set. For now,
we will consider only sets of this form. Second, instead of considering
whether $M$ is internally dense, we will examine the probability
of the stronger event of $M$ being a clique. Finally, let us pretend
all vertices in $M$ are at height $h$ from each other, so that the
probability of any edge inside $M$ occuring is $c^{-h}$. Note that
this last simplification is not so extreme: for any vertex $v\in M$,
$\frac{b-1}{b}\left|M\right|$ of the vertices in $M$ (that is, a
large fraction) are height exactly $h$ from $v$; of the remaining
$\frac{\left|M\right|}{b}$ vertices, most have height close to $h$.

Since there are ${b^{h} \choose 2}$ potential edges in $M,$ the
probability that $M$ is a clique is therefore at least 
\[
\left(c^{-h}\right)^{{b^{h} \choose 2}}\geq c^{-hb^{2h}}
\]

There are $\frac{n}{b^{h}}$ disjoint complete sets of height $h$,
so if we let $X$ denote the number of complete cliques of height
$h$, we have that
\begin{align}
E[X] & \geq\frac{n}{b^{h}}c^{-hb^{2h}}\nonumber \\
 & \approx\exp\left(\ln n-hb^{2h}\right)\label{eq:intuition}
\end{align}

(Of course, this is not really correct due to the simplifications
we have made, but turns out to be close enough to give the right asymptotic
value.) For the expected number of cliques to be growing with $n$,
we require the positive term in the exponential in \prettyref{eq:intuition}
to be growing faster than the negative term. Suppose the height is
given $h=a\frac{\ln\ln n}{\ln b}$ for some constant $a$, so that
the size of $M$ is $b^{h}=\left(\ln n\right)^{a}$. Then \prettyref{eq:intuition}
shows that the expected number of cliques will grow precisely when
$a<\frac{1}{2}$, the value of our cluster size threshold. 

Of course, the preceding justification considers only complete sets,
and ignores many details such as external sparseness. Many other types
of sets may exist, and it is natural to suspect they could form clusters
with sizes larger than $\left(\ln n\right)^{\frac{1}{2}+\epsilon}$.
Much of the work in proving part b) of \prettyref{thm:Thm1} is in
showing that this is not the case. 

Let us try to give intuition for why part b) of \prettyref{thm:Thm1}
should be true; that is, why clusters larger than $\left(\ln n\right)^{\frac{1}{2}+\epsilon}$
should not exist. As noted, the sets most likely to be internally
dense are those with small heights, because the probability of edges
is higher. In this simplified explanation, we will consider only sets
of this form. Let $S$ be a complete set of height $h=\left(\frac{1}{2}+\epsilon\right)\frac{\ln\ln n}{\ln b}$,
so that $S$ has $b^{h}=\left(\ln n\right)^{\frac{1}{2}+\epsilon}$
vertices and ${b^{h} \choose 2}\approx\left(\ln n\right)^{1+2\epsilon}$
possible edges. Of these edges, only $b^{\frac{h}{2}}{b^{\frac{h}{2}} \choose 2}=o\left(\ln n\right)$
have height $\frac{h}{2}$ or less. Ignoring these edges, which are
negligible in number, the remaining edges occur with probability at
most $c^{-\frac{h}{2}}$. This turns out to be small enough to show
that a.a.s. there are no complete sets like $S$ with more than $\ln n$
edges. 

That is, each complete set $S$ of height $h=\left(\frac{1}{2}+\epsilon\right)\frac{\ln\ln n}{\ln b}$
does not contain many edges. Given a set $M$, edges from $M\cap S$
to $M\setminus S$ occur with low probability, because the height
between these two sets is high. If the intersection $M\cap S$ is
large enough, then $M$ also cannot have many edges contained within
$S$, because $S$ itself does not have many edges. On the other hand,
if $M$ does not overlap any complete set $S$ significantly, then
$M\cap S$ is small enough to be ignored. In other words, if $M$
is large enough, then $M$ cannot be internally dense, and is therefore
not a cluster.

Although this explanation gives the intuition behind the core idea
of the proof of \prettyref{thm:Thm1}, the actual proof is much complicated,
requiring multiple steps to complete. 

Our result holds as well in the directed version of the CGA model
using the directed definition of clusters. This version of the theorem
is stated as follows:
\begin{thm}
\label{thm:Thm2-new}Let $G$ be a graph chosen according to the directed
CGA model. Then for all $0<\alpha,\beta\leq1$ and $\epsilon>0$:
\begin{description}
\item [{a}] Let $m^{\star}=\frac{\ln b}{\alpha\ln c}$. There a.a.s. exists
directed $\left(\alpha,\beta\right)$-clusters of size larger than
$\left(\ln n\right)^{\frac{1}{2}-\epsilon}$ in $G$. Moreover, there
exists a constant $\gamma>0$ such that for each $h$ satisfying $\log_{b}m^{\star}<h\leq\left(\frac{1}{2}-\epsilon\right)\frac{\ln\ln n}{\ln b}$,
there a.a.s. exists at least $\left(\ln n\right)^{\gamma b^{h}}$
complete directed $\left(\alpha,\beta\right)$-clusters of size $b^{h}$.
\item [{b}] There are a.a.s. no $\left(\alpha,\beta\right)$-clusters with
more than $\left(\ln n\right)^{\frac{1}{2}+\epsilon}$ vertices.
\end{description}
\end{thm}
\begin{rem*}
In fact, the directed version is easier to work with than the undirected
one, due to the added independence of having one potential edge in
each direction between each node pair. A different argument than the
one given here strengthens the number of clusters given in part a)
of \prettyref{thm:Thm2-new}: we can show that a.a.s. there exists
at least $\frac{n}{\ln n}$ complete directed $\left(\alpha,\beta\right)$-clusters
of size $b^{h}$.
\end{rem*}
We give the proof of \prettyref{thm:Thm1} but not \prettyref{thm:Thm2-new}.
The proof of \prettyref{thm:Thm2-new} is nearly identical to that
of \prettyref{thm:Thm1}, and may be given with a few straightforward
changes. The remainder of the paper will be organized as follows.
\prettyref{sec:Related-Work} lists related work. \prettyref{sec:The-Existence-of-1}
gives several results establishing the existence and non-existence
of clusters of size less than $\left(\ln n\right)^{\frac{1}{2}}$,
including a proof of part a) of \prettyref{thm:Thm1}. \prettyref{sec:The-Non-Existence-of}
gives a proof of part b) of \prettyref{thm:Thm1}.

\section{Related Work\label{sec:Related-Work}}

\subsection{Random Models for Real-World Networks}

Models for real-world networks have evolved in response to several
important empirical observations. One of these is the \emph{small-world}
\emph{effect}, the observation that network diameters are smaller
than one might expect \cite{Watts1998,Albert1999,Bollobas2004,Milgram1967}.
Watts and Strogatz \cite{Watts1998} and Kleinberg \cite{Kleinberg2000}
proposed models that add random edges to a regular network to reduce
the diameter. Another key property is the \emph{scale-free} property:
degree distributions tend to follow a power law.%
\footnote{That is, the proportion of nodes with degree $d$ is proportional
to $d^{-\zeta}$ for some constant $\zeta>0$.%
} This was first observed in the Internet graph \cite{Faloutsos1999},
and later in phone call graphs \cite{Abello1998} and the web graph
\cite{Barabasi2000}. \emph{Preferential attachment} models \cite{Barabasi1999,Bollobas2001b,Bollobas2004},
in which a network grows by adding new edges with a preference for
attachment to nodes with high degree, give one approach to explaining
this. Another approach is \emph{edge copying} \cite{Kleinberg1999,Kumar2000},
in which newly added vertices copy the edges of existing vertices.
Many other models exist (see e.g. \cite{Aiello2000,Skyrms2000,Li2003,Bonato2010a,Kim2010,Chojnacki2010},
among many others).

In 2005, Leskovek, Kleinberg, and Faloutsos \cite{Leskovec2005} observed
that average node degree increased polynomially as the network grows.
Prior to this, models had assumed a constant (or possibly logarithmically
increasing) average node degree. They proposed two models with densification:
the Community Guided Attachment (CGA) model used in this work, and
a ``forest fire'' model similar to edge copying. Another approach
for densifying models is based on Kronecker graphs \cite{Leskovec2005a,Mahdian2007}.
Lattanzi and Sivakumar \cite{Lattanzi2009} recently gave a model
based on affiliation networks, in which each node is affiliated with
some number of ``societies''.

\subsection{Clustering}

Detection of clusters is greatly important in sociology, biology,
and computer science. Fortunato \cite{Fortunato2010} gives a good
review of some of the many hundreds of published works on the topic.
Depending on the application, the definition of what constitutes a
cluster can vary greatly. Empirical studies show that clustering is
present in real-world networks \cite{Watts1998,Barrat2000,Leskovec2009},
and that these clusters often overlap \cite{Palla2005,Adams2007}:
that is, a single node in a real-world network may be part of multiple
clusters at once. 

Most popular approaches to clustering (e.g. \cite{Kernighan1970,Rattigan2007,Newman2004,Holme2003,Andreev2006})
do not allow overlapping clusters. The most popular approach to overlapping
communities is the clique percolation method \cite{Adamcsek2006,Farkas2007,Lehmann2008}.
In this method, two $k$-cliques \emph{overlap} if they share $k-1$
vertices. A $k$\emph{-clique community }(cluster) is the union of
a $k$-clique with all other $k$-cliques that overlap it. One problem
with this approach is that it is not clear initially which value of
$k$ should be chosen. Additionally, it presumes the existence of
many $k$-cliques, which may not be the case. Mishra, Schreiber, Stanton,
and Tarjan's \cite{Mishra2007} $\left(\alpha,\beta\right)$-clusters
(the clustering definition used in this work) avoids these problems
by instead parameterizing the fraction of edges that should be present
inside the cluster (the parameter $\beta$). Additionally, they introduce
the notion of external sparseness (the parameter $\alpha$). Other
approaches to overlapping clusters exist (e.g. \cite{Baumes2005,Evans2009,Zhang2007,Pinney2006}).

Investigations into the size of clusters in real-world networks show
that the tail of the cluster size distribution may follow a power
law \cite{Palla2005,Radicchi2004,Newman2004a}. In other words, the
relative sizes of the larger clusters in a network follow a certain
distribution; unlike the current work, no observation is made about
size relative to the overall size of the network. Recently, Leskovek,
Lang, Dasgupta, and Mahoney \cite{Leskovec2009} found empirical evidence
for the existence of a size threshold for the ``best'' clusters
in the network: beyond that size, the quality of clusters declines.
This is also discussed in \prettyref{sec:Introduction}.

Despite the importance of clustering and the proliferation of random
models for real-world networks, we are aware of no work that studies
clustering analytically in any random model for real-world networks.

\section{The Existence of Small Clusters\label{sec:The-Existence-of-1}}

To prove the existence of clusters, we must establish that both external
sparseness and internal denseness occur. We begin with external sparseness.
Let $h^{\star}=\left(\frac{1}{2}-\epsilon\right)\frac{\ln\ln n}{\ln b}$.
This value is important because the number of nodes in a complete
set of height $h^{\star}$ is equal to $b^{h^{\star}}=\left(\ln n\right)^{\frac{1}{2}-\epsilon}$,
the bound in \prettyref{thm:Thm1} a). Let $M$ be a set with height
$h\leq h^{\star}$ and size $m$. Recall that $S\left(M\right)$ is
the complete set of height $h$ containing $M$, and $S\left(M,h^{\star}\right)$
is the complete set of height $h^{\star}$ containing $M$. For $M$
to be externally sparse, we require three events to occur, defined
as follows: 
\begin{itemize}
\item \textbf{$\mathbf{E_{1}}\left(M\right)$}: The vertices of $S(M)\setminus M$
must satisfy the external sparseness property with respect to $M$.
That is, $\forall u\in S\setminus M$, $e(u,M)\leq\alpha m$.
\item \textbf{$\mathbf{E_{2}}\left(M\right)$}: The vertices of $S\left(M,h^{\star}\right)\setminus S(M)$
must satisfy the external sparseness property with respect to $M$.
That is, $\forall u\in S\left(M,h^{\star}\right)\setminus S(M)$,
$e(u,M)\leq\alpha m$.
\item \textbf{$\mathbf{E_{3}}\left(M\right)$}: The vertices of $G\setminus S\left(M,h^{\star}\right)$
must satisfy the external sparseness property with respect to $M$.
That is, $\forall u\in G\setminus S\left(M,h^{\star}\right)$, $e(u,M)\leq\alpha m$.
\end{itemize}
When the set $M$ is clear, we will sometimes denote these events
as $\mathbf{E_{1}}$, $\mathbf{E_{2}}$, and $\mathbf{E_{3}}$, without
the parenthetical argument. Together, $\mathbf{E_{1}}\cap\mathbf{E_{2}}\cap\mathbf{E_{3}}$
forms the event that $M$ is externally sparse. This division will
be used to show in steps when these different parts of the external
sparseness property occur. In particular, this lessens the problem
of dependence between sets. To see this, let $M_{1}$ and $M_{2}$
be sets such that $S\left(M_{1},h^{\star}\right)$ and $S\left(M_{2},h^{\star}\right)$
do not intersect. Because these sets do not intersect, the events
\textbf{$\mathbf{E_{1}}\left(M_{1}\right)$}, \textrm{$\mathbf{E_{1}}\left(M_{2}\right)$,
}\textbf{$\mathbf{E_{2}}\left(M_{1}\right)$}, and \textbf{$\mathbf{E_{2}}\left(M_{2}\right)$}
are all independent; the only events with dependence between each
other are \textbf{$\mathbf{E_{3}}\left(M_{1}\right)$ }and \textbf{$\mathbf{E_{3}}\left(M_{2}\right)$}. 

The following lemma establishes when $\mathbf{E_{2}}$ can occur:
\begin{lem}
\label{lem:E_1_and__E2-1}Let $m^{\star}=\frac{\ln b}{\alpha\ln c}$
and $h^{\star}=\left(\frac{1}{2}-\epsilon\right)\frac{\ln\ln n}{\ln b}$.
Then:
\begin{description}
\item [{a}] There are a.a.s. no externally sparse sets of size smaller
than $m^{\star}$.
\item [{b}] If $M$ is a set of size $m>m^{\star}$ and height $h\leq h^{\star}$,
then there exists a constant $a>0$ such that $\Pr\left(\mathbf{E_{2}}\left(M\right)\right)>a.$
\end{description}
\end{lem}
The proof of \prettyref{lem:E_1_and__E2-1} may be found in the appendix;
a) follows from a first moment argument, and b) comes from the application
of a concentration bound. 

Because of the dependence problem, dealing with $\mathbf{E_{3}}$
is more complicated. To give intuition, consider the following scenario.
Suppose $G$ is partitioned into complete sets of height $h^{\star}$,
and from each of these sets we choose at most a single subset of size
$m$. Call this resulting collection of subsets $\mathcal{M}$. Any
single one of those subsets is likely to have $\mathbf{E_{3}}$, because
the probability of edges from that set to any of the other sets is
small. In fact, by dealing with dependence appropriately, we can show
that a.a.s. at least 

\begin{equation}
\min\left(\left|\mathcal{M}\right|,\left(\ln n\right)^{\frac{\alpha\ln c}{4\ln b}\left(m-m^{\star}\right)}\right)\label{eq:size_of_M}
\end{equation}
of the sets in $\mathcal{M}$ have $\mathbf{E_{3}}$ simultaneously.
The proof of \prettyref{eq:size_of_M} may be found in the appendix.

We use \prettyref{eq:size_of_M} and \prettyref{lem:E_1_and__E2-1}
to prove the following corollary:
\begin{cor}
\label{cor:Constant-size-existence-1}Let $m^{\star}=\frac{\ln b}{\alpha\ln c}$,
and let $h=\Theta(1)$ be a fixed, constant value. Then:
\begin{description}
\item [{a}] For $m<m^{\star}$, there are a.a.s. no clusters of size $m$. 
\item [{b}] For each $m$ such that $m^{\star}<m\leq b^{h}$, there are
a.a.s. at least \textrm{\textup{$\left(\ln n\right)^{\frac{\alpha\ln c}{4\ln b}\left(m-m^{\star}\right)}$}}
clusters of size $m$ and height $h$.
\end{description}
\end{cor}
Note that Corollary \ref{cor:Constant-size-existence-1} establishes
a sharp size bound on the existence of small, constant-sized clusters.
We give the proof of \prettyref{cor:Constant-size-existence-1}:
\begin{proof}
Because a set must be externally sparse to be a cluster, part a) follows
directly from \prettyref{lem:E_1_and__E2-1}a). To show part b), let
$M$ be a complete set of height $h$ and size $m>m^{\star}$, and
let $\mathbf{D}$ denote the event that $M$ is internally dense.
Because there are only a constant number of vertices in $M$, $M$
has $\mathbf{D}$ with constant probability. Similarly, there are
only a constant number of vertices in $S(M)\setminus M$, so $M$
has $\mathbf{E_{1}}$ with constant probability. By \prettyref{lem:E_1_and__E2-1}
b), $M$ has $\mathbf{E_{2}}$ with at least constant probability.
Hence, there is a constant $q>0$ such that $\Pr\left(\mathbf{D}\cap\mathbf{E_{1}}\cap\mathbf{E_{2}}\right)>q$. 

Now, partition $G$ into $n/b^{h^{\star}}$ sets of height $h^{\star}=\left(\frac{1}{2}-\epsilon\right)\frac{\ln\ln n}{\ln b}$,
and from each set choose a complete set of height $h$. Because each
set lies in a different complete set of height $h^{\star}$, each
set has $\mathbf{D}\cap\mathbf{E_{1}}\cap\mathbf{E_{2}}$ with probability
at least $a$ independently. Let those sets that have $\mathbf{D}\cap\mathbf{E_{1}}\cap\mathbf{E_{2}}$
form the collection of sets $\mathcal{M}$. The expected size of $\mathcal{M}$
is $qn/b^{h^{\star}}=\Omega\left(n^{3/4}\right)$, and a concentration
bound (see appendix) shows that a.a.s. $\left|\mathcal{M}\right|>n^{3/4}$.
Applying \prettyref{eq:size_of_M}, we see that at least \textrm{$\left(\ln n\right)^{\frac{\alpha\ln c}{4\ln b}\left(m-m^{\star}\right)}$}
of the sets in $\mathcal{M}$ are clusters.
\end{proof}
We are now ready to show how \prettyref{eq:size_of_M} proves part
a) of \prettyref{thm:Thm1}, which we now restate:
\begin{quotation}
\emph{Let $m^{\star}=\frac{\ln b}{\alpha\ln c}$. There a.a.s. exists
clusters of size larger than $\left(\ln n\right)^{\frac{1}{2}-\epsilon}$
in $G$. Moreover, there exists a constant $\gamma=\gamma\left(\alpha,b,c\right)>0$
such that for each $h$ satisfying $\log_{b}m^{\star}<h\leq\left(\frac{1}{2}-\epsilon\right)\frac{\ln\ln n}{\ln b}$,
there a.a.s. exists at least $\left(\ln n\right)^{\gamma b^{h}}$
complete clusters of size $b^{h}$.}\end{quotation}
\begin{proof}
Note that the second part of the theorem statement implies the first,
because a complete set of height $h^{\star}=\left(\frac{1}{2}-\epsilon\right)\frac{\ln\ln n}{\ln b}$
has $b^{h^{\star}}=\left(\ln n\right)^{\frac{1}{2}-\epsilon}$ vertices.
Let $M$ be a complete set of height $h$. We will first examine the
event $\mathbf{D}$ that $M$ is internally dense. 

In fact, let us consider the stronger event of $M$ being a clique.
Each potential edge in $M$ occurs with probability at least $c^{-h}$.
so the probability of all ${b^{h} \choose 2}\leq b^{2h}$ edges occuring
is at least $c^{-hb^{2h}}$. Hence, $\Pr\left(\mathbf{D}\right)\geq\exp\left(-\ln\left(c\right)hb^{2h}\right)$. 

Now, partition $G$ into complete sets of height $h^{\star}=\left(\frac{1}{2}-\epsilon\right)\frac{\ln\ln n}{\ln b}$.
From each such set, choose a single complete set of height $h$, and
let these $n/b^{h^{\star}}$ disjoint sets form the family $\mathcal{A}$.
Because these sets are complete, they automatically have $\mathbf{E_{1}}$.
Because they do not overlap and lie in different sets of height $h^{\star}$,
they will have $\mathbf{D}$ and $\mathbf{E_{2}}$ independently.
\prettyref{lem:E_1_and__E2-1} b) implies they have $\mathbf{E_{2}}$
with probability at least $a$ for some positive constant $a$. Hence,
for every $M\in\mathcal{A}$, we have that
\begin{align}
\Pr\left(\mathbf{D}\cap\mathbf{E_{1}}\cap\mathbf{E_{2}}\right) & \geq\Pr\left(\mathbf{D}\right)\Pr\left(\mathbf{E_{1}}\right)\Pr\left(\mathbf{E_{2}}\right)\nonumber \\
 & \geq a\exp\left(-\ln\left(c\right)hb^{2h}\right)\label{eq:D_E1_E2_lower_bd-1}
\end{align}

Now, let $X$ denote the number of sets in $\mathcal{A}$ with $\mathbf{D}\cap\mathbf{E_{1}}\cap\mathbf{E_{2}}$.
A fairly straightforward calculation (found in the appendix) shows
that $E[X]$ is asymptotically larger than $\frac{n}{\ln n}$, and
moreover that \textrm{$\Pr\left(X<\frac{n}{\ln n}\right)\leq\exp\left(n^{-1/3}\right)$.}
In other words, a.a.s. there are at least $\frac{n}{\ln n}$ sets
in $\mathcal{A}$ with $\mathbf{D}\cap\mathbf{E_{1}}\cap\mathbf{E_{2}}$.
Let $\mathcal{M}\subset\mathcal{A}$ be this sub-family of sets. \prettyref{eq:size_of_M}
implies that at least \textrm{$\left(\ln n\right)^{\frac{\alpha\ln c}{4\ln b}\left(b^{h}-m^{\star}\right)}$
of the sets in $\mathcal{M}$ also have $\mathbf{E_{3}}$, and therefore
are clusters.} Recall that $m^{\star}=\frac{\ln b}{\alpha\ln c}$,
and let $h_{min}$ be the minimum integral value of $h$ such that
$b^{h}>m^{\star}$; setting 
\[
\gamma=\frac{\alpha\ln c}{4\ln b}\frac{\left(b^{h_{min}}-m^{\star}\right)}{b^{h_{min}}}
\]
guarantees that $\left(\ln n\right)^{\gamma b^{h}}\leq\left(\ln n\right)^{\frac{\alpha\ln c}{4\ln b}\left(b^{h}-m^{\star}\right)}$,
and hence there are a.a.s. at least $\left(\ln n\right)^{\gamma b^{h}}$
clusters of height $h$ in $G$.
\end{proof}

\section{The Non-Existence of Large Clusters\label{sec:The-Non-Existence-of}}

The goal of this section is to prove part b) of \prettyref{thm:Thm1}:
that, for all $\epsilon>0$, there are no clusters with more than
$\left(\ln n\right)^{\frac{1}{2}+\epsilon}$ vertices. 

The intuition of the proof may be outlined as follows. A set of a
given size is more likely to form a cluster if the height of that
set is small. More generally, a set is more likely to be a cluster
if a large subset of the set has a small height, so that edges within
that subset are more common. We will concentrate on ruling out these
types of clusters. 

We use the term \emph{thick }to capture this notion of a set with
small height containing many vertices. We will consider the following
two types of thick sets:
\begin{defn*}
$\,$
\begin{itemize}
\item A\emph{ short $\epsilon$-thick set} is a set with height at most
$h_{\epsilon}=\left(\frac{1}{2}+\epsilon\right)\frac{\ln\ln n}{\ln b}$
and containing at least $\left(\ln n\right)^{\frac{1}{2}+\frac{\epsilon}{3}}$
vertices.
\item A \emph{tall $\epsilon$-thick set }is a set with height at most $\frac{\left(\ln n\right)^{\frac{1}{2}}}{\ln b}$
and containing at least $\left(\ln n\right)^{\frac{1}{2}+\frac{\epsilon}{2}}$
vertices. 
\end{itemize}
\end{defn*}
Note that the height $h_{\epsilon}$ is the height of a complete set
with $\left(\ln n\right)^{\frac{1}{2}+\epsilon}$ nodes, as in \prettyref{thm:Thm1}
b).

Let $Q$ be a set of size $q\geq\left(\ln n\right)^{\frac{1}{2}+\epsilon}$.
There are two cases: either $Q$ contains no tall $\epsilon$-thick
sets, or it contains at least one such set as a subset. Existence
of clusters in the former case is easy to rule out: given any vertex
$v\in Q$, we must have at least $\beta q$ edges from $v$ to other
vertices in $Q$ for $Q$ to be a cluster. Since there is no tall
$\epsilon$-thick set, at most $\left(\ln n\right)^{\frac{1}{2}+\frac{\epsilon}{2}}=o(q)$
of these vertices are ``close''---that is, within height $\frac{\left(\ln n\right)^{\frac{1}{2}}}{\ln b}$
of $v$. Even assuming all the edges from $v$ to the close vertices
exist, we must have at least $\frac{\beta}{2}q$ edges from $v$ to
the far vertices in $Q$, which occur with low probability.

The latter case is harder. Let us write $Q=T\cup R$, where $T$ is
a tall $\epsilon$-thick set, and $R$ contains the other vertices.
Given $v\in T$, we must also rule out the possibility of there being
many edges between $v$ and $T$, which occur with much higher probability
than edges between $v$ and $R$. 

This is done by repeating the argument used above for tall $\epsilon$-thick
sets on short $\epsilon$-thick sets. $T$ will either contain a short
$\epsilon$-thick set as a subset, or it will not. Again, the latter
case is more easy to deal with, because most of the required edges
for $Q$ to form a cluster are of height $h_{\epsilon}=\left(\frac{1}{2}+\epsilon\right)\frac{\ln\ln n}{\ln b}$
or more, and so are unlikely to occur. In the former case, we partition
$T=M\cup K$ so that $M$ is the subset of height $h_{\epsilon}$
with at least $\left(\ln n\right)^{\frac{1}{2}+\frac{\epsilon}{3}}$
vertices, and $K$ are the remaining vertices. We now have $Q=M\cup K\cup R$;
a simplified form of the argument for this case is as follows. We
will show that there is some $v\in M$ such that $e(v,M)\leq\frac{\beta}{2}m$,
$e(v,K)\leq\frac{\beta}{2}k$, and $e(v,R)\leq\frac{\beta}{2}r$,
together implying that $e(v,Q)\leq\frac{\beta}{2}q$.

However, some care is needed to make the asymptotics of this argument
work. There are more than ${n \choose q}$ choices for the set $Q$;
if we divide $Q=M\cup K\cup R$ for each such set $Q$, there are
far too many choices of the set $M$ for a first moment bound to show
directly that a.a.s. all sets $M$ have the property we desire. For
that reason, the argument will instead be given in the reverse order:
first, we will show that a.a.s. all short $\epsilon$-thick sets have
the property we desire. This will be used to show that tall $\epsilon$-thick
sets also have a desired property. Finally, the result on tall $\epsilon$-thick
sets will be used to prove the clustering threshold for all sets of
size at least $\left(\ln n\right)^{\frac{1}{2}+\epsilon}$ in general.

We point out the need to make similar arguments twice, once with a
height of $h_{\epsilon}=\left(\frac{1}{2}+\epsilon\right)\frac{\ln\ln n}{\ln b}$,
and once with a height of $\frac{\left(\ln n\right)^{\frac{1}{2}}}{\ln b}$.
It is possible to show directly that a.a.s. all short $\epsilon$-thick
sets are not clusters, but the asymptotics of this argument will not
work with a set of larger size. By splitting sets of height $\frac{\left(\ln n\right)^{\frac{1}{2}}}{\ln b}$
up into both sets containing short $\epsilon$-thick sets and those
that do not, this in turn implies that a tall $\epsilon$-thick sets
are also a.a.s. not clusters. Again, the asymptotics of this argument
will not work for a sets of larger size. Finally, we consider all
sets with at least $\left(\ln n\right)^{\frac{1}{2}+\epsilon}$ vertices.
By splitting the class of such sets up into both those that contain
tall $\epsilon$-thick sets and those that do not, we may finally
show no clusters with at least $\left(\ln n\right)^{\frac{1}{2}+\epsilon}$
vertices exist.

We begin by precisely stating the property of short $\epsilon$-thick
sets that we are interested in, and showing when it occurs: 
\begin{lem}
\label{lem:thick-cluster-upper-bd}For all $\epsilon$ such that $\frac{\ln c}{4\ln b}>\epsilon>0$,
a.a.s. for each short $\epsilon$-thick set $M$, there exists a set
$M_{1}\subset M$ such that $\left|M_{1}\right|\geq\frac{3}{4}\left|M\right|$
and $\forall v\in M_{1}$, $e(v,M)\leq\frac{\beta}{4}\left|M\right|$.
\end{lem}
The proof of \prettyref{lem:thick-cluster-upper-bd} may be found
in the appendix. \prettyref{lem:thick-cluster-upper-bd} implies that
there are a.a.s. no short $\epsilon$-thick clusters. The stronger
notion used here (that \emph{many} vertices in $M$ have at most $\frac{\beta}{4}\left|M\right|$
edges) is necessary for later steps of the proof.

The next lemma is the first step in showing tall $\epsilon$-thick
sets (of height $\frac{\left(\ln n\right)^{\frac{1}{2}}}{\ln b}$)
are not clusters: 
\begin{lem}
\label{lem:L7-NoC_K sets with many bad vertices}For all $\epsilon>0$,
a.a.s. for every tall $\epsilon$-thick set $T=M\cup K$, where $M$
is a short $\epsilon$-thick set and $K$ is a set with $\left|K\right|\geq\frac{\beta}{4}\left|M\right|$
such that $K$ does not intersect $S\left(M,h_{\epsilon}\right)$,
the complete set of height $h_{\epsilon}=\left(\frac{1}{2}+\epsilon\right)\frac{\ln\ln n}{\ln b}$,
there exists a set $M_{2}\subset M$ such that $\left|M_{2}\right|>\frac{3}{4}\left|M\right|$
and $\forall v\in M$, $e\left(v,K\right)\leq\frac{\beta}{2}\left|K\right|$.
\end{lem}
The proof of \prettyref{lem:L7-NoC_K sets with many bad vertices}
may be found in the appendix. Intuitively, \prettyref{lem:thick-cluster-upper-bd}
and \prettyref{lem:L7-NoC_K sets with many bad vertices} go together
as follows: let $T=M\cup K$ be a tall $\epsilon$-thick set containing
a short $\epsilon$-thick set $M$ as well as some other vertices
$K$. \prettyref{lem:thick-cluster-upper-bd} implies that there are
not enough edges within $M$ for $T$ to be internally dense, and
\prettyref{lem:L7-NoC_K sets with many bad vertices} implies similarly
that the edges from $M$ to $K$ are not sufficient for $T$ to be
internally dense. Taken together, this shows that a.a.s. there are
no tall $\epsilon$-thick clusters:
\begin{lem}
\label{lem:Tall_clusters_have_many_bad_vertices}For all $\epsilon$
such that $\frac{\ln c}{4\ln b}>\epsilon>0$, a.a.s. for every tall
$\epsilon$-thick set $T$, there exists a set $T^{\prime}\subset T$
such that $\left|T^{\prime}\right|>\left(\ln n\right)^{\frac{1}{2}+\frac{\epsilon}{4}}$
and $\forall v\in T^{\prime}$, $e\left(v,T\right)\leq\frac{\beta}{2}\left|T\right|$.
\end{lem}
Because it is required for the proof of later steps, \prettyref{lem:Tall_clusters_have_many_bad_vertices}
shows a property stronger than that of not being a cluster. The proof
of \prettyref{lem:Tall_clusters_have_many_bad_vertices} is found
in the appendix, but is sketched as follows:
\begin{proof}[Proof Sketch]
First, suppose $T$ contains a short $\epsilon$-thick set $M$.
Adding vertices in $S\left(M,h_{\epsilon}\right)\setminus M$ to $M$
will preserve the property that $M$ is a short $\epsilon$-thick
set, so we may assume that $M$ is maximal. In other words, we may
assume the set $T\setminus M$ does not intersect the compelte set
of height $h_{\epsilon}$, $S\left(M,h_{\epsilon}\right)$. Set $K=T\setminus M$,
and Let $m=\left|M\right|$ and $k=\left|K\right|$. By \prettyref{lem:thick-cluster-upper-bd},
there is a set $M_{1}\subset M$ such that $\left|M_{1}\right|\geq\frac{3}{4}m$
and $\forall v\in M_{1}$, $e\left(v,M\right)\leq\frac{\beta}{4}m$.
If $k<\frac{\beta}{4}m$, then even if every edge from $v$ to $K$
exists, we will still have $e\left(v,M\cup K\right)\leq\frac{\beta}{4}m+k\leq\frac{\beta}{2}m\leq\frac{\beta}{2}t$
for every $v\in M_{1}$, and hence may choose $T^{\prime}=M_{1}$.
If $k\geq\frac{\beta m}{4}$, then by \prettyref{lem:L7-NoC_K sets with many bad vertices},
there is some set $M_{2}\subset M$ such that $\left|M_{2}\right|>\frac{3}{4}m$
and $\forall v\in M_{2}$, $e\left(v,K\right)\leq\frac{\beta}{2}k$.
Taking $T^{\prime}=M_{1}\cap M_{2}$, it follows that $\left|T^{\prime}\right|\geq\frac{m}{2}$
and $e\left(v,M\cup K\right)\leq\frac{\beta}{2}m+\frac{\beta}{2}k=\frac{\beta}{2}t$
for every $v\in T^{\prime}$. 

In the case that $T$ contains no short $\epsilon$-thick clusters,
every $v\in T$ has at most $o\left(t\right)$ vertices that are close
to $v$, and the rest are far away and therefore these edges occur
with low probability. It is thus not hard to show with a first moment
argument that $T$ has the desired property.
\end{proof}
We are now ready to prove part b) of \prettyref{thm:Thm1}, which
we now restate:
\begin{quotation}
\emph{For all $\epsilon>0$, there are a.a.s. no clusters with more
than $\left(\ln n\right)^{\frac{1}{2}+\epsilon}$ vertices.}
\end{quotation}
The proof will occur in two steps. First, we give the following lemma:
\begin{lem}
\label{lem:No_clusters_containing_tall_sets}For all $\epsilon$ such
that $\frac{\ln c}{4\ln b}>\epsilon>0$, there are a.a.s. no clusters
$Q=T\cup R$ of size at least $\left(\ln n\right)^{\frac{1}{2}+\epsilon}$,
where $T$ is a tall $\epsilon$-thick set.
\end{lem}
The proof of \prettyref{lem:No_clusters_containing_tall_sets}, achieved
using the first moment method, is found in the appendix. This lemma
rules out the most likely type of potential cluster, leaving only
sets with at least $\left(\ln n\right)^{\frac{1}{2}+\epsilon}$ vertices
that contain no tall $\epsilon$-thick subsets. We now prove part
b) of \prettyref{thm:Thm1} by considering this case.
\begin{proof}
Clearly if the theorem holds for all $\epsilon$ such that $\frac{\ln c}{4\ln b}>\epsilon>0$,
then it will hold for all $\epsilon>0$, so we may assume $\epsilon<\frac{\ln c}{4\ln b}$.
Let $Q$ be a potential cluster of size $q\geq\left(\ln n\right)^{\frac{1}{2}+\epsilon}$.
By \prettyref{lem:No_clusters_containing_tall_sets}, we may assume
$Q$ contains no tall $\epsilon$-thick sets; that is, $Q$ contains
no sets of height at most $h=\frac{\left(\ln n\right)^{\frac{1}{2}}}{\ln b}$
with at least $z=\left(\ln n\right)^{\frac{1}{2}+\frac{\epsilon}{2}}$
vertices. Hence, if we subdivide the vertices of $G$ into $\frac{n}{b^{h}}$
sets of height $h$, $Q$ must have fewer than $z$ vertices inside
each set. This implies that for all $v\in Q$, at least $q-z>\left(\ln n\right)^{\frac{1}{2}+\frac{\epsilon}{2}}$
vertices in $Q$ are at height $h$ or more from $v$; for $Q$ to
be a cluster, at least $\beta q-z>\frac{\beta}{2}q$ edges from $v$
to these distant vertices must exist. Let $X_{v}$ denote the number
of edges from $v$ to $Q$ that have height more than $h$, and let
$X_{Q}=\frac{1}{2}\sum_{v\in Q}X_{v}$ be the total number of such
edges in $Q$. If $X_{Q}\leq\frac{\beta}{4}q^{2}$, then it follows
that at least one vertex $v$ has $X_{v}\leq\frac{\beta}{2}q$, implying
that $Q$ is not a cluster.

Since there are less than ${q \choose 2}$ edges in $Q$ with height
more than $h$, each occurring with probability at most $c^{-h}$,
$X_{Q}$ is stochastically dominated by the random variable $\mbox{Bin}\left({q \choose 2},c^{-h}\right)$,
which has expected value ${q \choose 2}c^{-h}=o\left(q^{2}\right)$.
A concentration bound (see the appendix) therefore shows that 
\begin{align}
\Pr\left(X_{Q}\geq\frac{\beta}{4}q^{2}\right) & \leq\exp\left(-\frac{\beta\ln c}{8}q^{2}h\right).\label{eq:thm1b-bound-on-X_Q}
\end{align}

Now let $X$ be the number of sets $Q$ of size at least $\left(\ln n\right)^{\frac{1}{2}+\epsilon}$
such that $X_{Q}\geq\frac{\beta}{4}\left|Q\right|^{2}$. As noted
above, it suffices to show that a.a.s. $X=0$. Since there are less
than ${n \choose q}\leq\exp\left(q\ln n\right)$ choices of size $q$
for the set $Q$, we have

\begin{align*}
E[X] & \leq\sum_{q\geq\left(\ln n\right)^{\frac{1}{2}+\epsilon}}{n \choose q}\Pr\left(X_{Q}\geq\frac{\beta}{4}q^{2}\right)\leq\sum_{q\geq\left(\ln n\right)^{\frac{1}{2}+\epsilon}}\exp\left(q\ln n-\frac{\beta\ln c}{8}q^{2}h\right)
\end{align*}

Since $h=\frac{\left(\ln n\right)^{\frac{1}{2}}}{\ln b}$ and $q\geq\left(\ln n\right)^{\frac{1}{2}+\epsilon}$,
it follows that $q\ln n=o\left(q^{2}h\right)$ and the term in the
exponential is negative. Hence, it is maximized when $q$ is minimized;
that is, when $q=\left(\ln n\right)^{\frac{1}{2}+\epsilon}$. The
above becomes 
\begin{align*}
E[X] & \leq\sum_{q\geq\left(\ln n\right)^{\frac{1}{2}+\epsilon}}\exp\left(\left(\ln n\right)^{\frac{3}{2}+\epsilon}-\frac{\beta\ln c}{8\ln b}\left(\ln n\right)^{\frac{3}{2}+2\epsilon}\right)\\
 & \leq n\:\exp\left(-\frac{\beta\ln c}{10\ln b}\left(\ln n\right)^{\frac{3}{2}+2\epsilon}\right)
\end{align*}
which tends to zero as $n\rightarrow\infty$. By the first moment
method, this implies that a.a.s. $X=0$.
\end{proof}

\section{Conclusions\label{sec:Conclusions}}

This work deals with the question: \emph{what form do clusters take
in real-world networks?} In this case, we considered the existence
of $\left(\alpha,\beta\right)$-clusters in the CGA real-world random
network model. We showed the existence of a size threshold of $\left(\ln n\right)^{\frac{1}{2}}$
for the existence of such clusters. As noted in \prettyref{sec:Introduction},
the CGA model captures many of the properties observed in real-world
networks. $\left(\alpha,\beta\right)$-clusters capture a particular
notion of clustering in real-world networks, in which clusters are
denser than their surrounding neighbourhood, and in which clusters
may overlap. Therefore, the choice of model and clustering definition
seem valid for approaching the motivating question. 

Thus, it is interesting to ask the extent to which the clustering
threshold of $\left(\ln n\right)^{\frac{1}{2}}$ extends beyond the
CGA model to random models for real-world networks in general. Can
a threshold for $\left(\alpha,\beta\right)$-clusters be observed
in real-world data? Can other models for real-world networks be shown
to have thresholds for cluster size? Do such size thresholds exist
for other notions of clustering?

One goal of this work was to achieve our result analytically. Many
real-world random network models are prohibitively hard to analyze,
so simulation is often needed to establish the existence of desirable
properties. We have avoided this approach and concentrated on analytic
results.

Turning now to our result, an open question is whether or not a sharp
threshold for $\left(\alpha,\beta\right)$-clusters exists: we have
not addressed the existence of $\left(\alpha,\beta\right)$-clusters
in the range $\left(\ln n\right)^{\frac{1}{2}\pm f(n)}$, where $f(n)=o\left(1\right)$.
Part a) of \prettyref{thm:Thm1} shows the existence of \emph{complete}
clusters at each height less than $\left(\frac{1}{2}-\epsilon\right)\frac{\ln\ln n}{\ln b}$,
but a more complete treatment might consider the existence of other
(i.e. non-complete) clusters of this size.

\section{Appendix}

This section gives the full proof of the results in this work. We
begin by introducing some probabilistic tools that will be required.

\subsection{Probabilistic Tools\label{sec:Probabilistic-Tools}}

We aim to characterize the asymptotic behaviour of $G$ as the number
of vertices $n$ increases. Since $G$ is only defined when $n$ is
a power of $b$, it is more correct to let $H$, the height of $G$,
increase, and take $n=b^{H}$. However, little clarity is lost when
taking asymptotics in relation to $n$.

The binomial random variable given by the number of successes over
$r$ independent trials, each succeeding with probability $p$, is
denoted $\mbox{Bin}\left(r,p\right)$.

The main probabilistic idea used is that of the \emph{first moment
method}: Let $X$ be a non-negative random variable that takes integral
values, with expected value $E[X]$. If $E[X]=o(1)$, then by Markov's
Inequality, $\Pr\left(X\geq1\right)=o(1)$. In other words, a.a.s.
$X=0$. This technique will be used repeatedly to establish that events
a.a.s. do not occur.

The following lemma gives a bound on the upper tail of a Binomial
random variable:
\begin{lem}
\label{lem:4-prob-tool}Let $X=\mbox{Bin}(n,p)$ be a binomial random
variable. Let $t>1$ and $1\leq s=\left\lceil tpn\right\rceil \leq n-1$.
Then

\[
\Pr\left(X\geq tpn\right)<\frac{t}{t-1}{n \choose s}p^{s}\left(1-p\right)^{n-s}
\]

\end{lem}
A proof may be found in \cite{Bollobas2001a}. We will use a simplified
form of this. Since ${n \choose s}\leq\left(\frac{ne}{s}\right)^{s}$
(which follows from Stirling's approximation), \prettyref{lem:4-prob-tool}
implies that if $s\geq2pn$, then

\begin{align}
\Pr\left(X\geq s\right) & \leq2{n \choose s}p^{s}\nonumber \\
 & \leq2\left(\frac{nep}{s}\right)^{s}\label{eq:L4-ptool-simple-intermediate}\\
 & \leq2\exp\left(s\left(\ln n+1-\ln s+\ln p\right)\right)\label{eq:L4-prob-tool-simple}
\end{align}

The final tool we will make use of is a Chernoff-type bound from Janson
\cite{Janson2002}, which gives concentration bounds on a sum of independent
Bernoulli random variables.
\begin{lem}
\label{lem:Bollobas-prob-tool}Let $X=X_{1}+...+X_{k}$, where the
$X_{i}$ are independent Bernoulli random variables with $\Pr\left(X_{i}=1\right)=p_{i}$.
Let $\mu=E[X]=\sum p_{i}$. Then for $t\geq0$ we have: 

\[
\Pr\left(X\geq\mu+t\right)\leq\exp\left(-\frac{t^{2}}{2(\mu+t/3)}\right)
\]

and 

\[
\Pr\left(X\leq\mu-t\right)\leq\exp\left(-\frac{t^{2}}{2\mu}\right).
\]

\end{lem}
A proof may be found in \cite{Janson2002}.

\subsection{Proofs\label{sec:Proofs}}

\theoremstyle{theorem}
\newtheorem*{lem3}{Lemma 3}
\begin{lem3} 
Let $m^{\star}=\frac{\ln b}{\alpha\ln c}$ and $h^{\star}=\left(\frac{1}{2}-\epsilon\right)\frac{\ln\ln n}{\ln b}$. Then: \begin{description} \item [{a}] There are a.a.s. no externally sparse sets of size smaller than $m^{\star}$. \item [{b}] If $M$ is a set of size $m>m^{\star}$ and height $h\leq h^{\star}$, then there exists a constant $a>0$ such that  \[ \Pr\left(\mathbf{E_{2}}\left(M\right)\right)>a. \]
\end{description} \end{lem3}
\begin{proof}
To prove $a)$, first suppose $M$ is a set of size $m<m^{\star}$.
The event $\mathbf{E}\left(M\right)$ that $M$ is externally sparse
holds if there does not exist any vertex $v\in V\left(G\right)\setminus M$
such that $e\left(v,M\right)>\alpha m$. Since $m=O\left(1\right)$,
we have $\left|V\left(G\right)\setminus M\right|\geq\frac{n}{2}$
for large enough $n$. The probability of an edge from $v$ to a vertex
in $M$ is at least $p=c^{-\log_{b}n}=n^{-\log_{b}c}$. It follows
that $e\left(v,M\right)$ stochastically dominates the random variable
$\mbox{Bin}\left(m,p\right)$, and hence 
\begin{equation}
\Pr\left(e\left(v,M\right)>\alpha m\right)>p^{\alpha m}.\label{eq:PrlowerBd-1}
\end{equation}
$\mathbf{E}\left(M\right)$ will hold only if all of the at least
$\frac{n}{2}$ vertices in $V\left(G\right)\setminus M$ have $\alpha m$
or fewer links, so we have

\begin{align}
\Pr\left(\mathbf{E}\left(M\right)\right) & \leq\left(1-p^{\alpha m}\right)^{\frac{n}{2}}\nonumber \\
 & \leq\exp\left(-\frac{1}{2}np^{\alpha m}\right)\nonumber \\
 & =\exp\left(-\frac{1}{2}n^{1-\alpha m\log_{b}c}\right).\label{eq:Pr_lemma_upper_bd_new}
\end{align}
Because $m<m^{\star}$, $n^{1-\alpha m\log_{b}c}$ goes to zero as
$n$ increases, so this probability is exponentially small.

Now, we wish to show that a.a.s. for every set $M$ of size less than
$m^{\star}$, $\mathbf{E}(M)$ does not hold. Let \textbf{$X$ }be
the number of clusters of size smaller than $m^{\star}$; since there
are at most ${n \choose m}$ sets of size $m$, it follows from \prettyref{eq:Pr_lemma_upper_bd_new}
that 

\begin{equation}
E[X]\leq\sum_{m<m^{\star}}{n \choose m}\exp\left(-\frac{1}{2}n^{1-\alpha m\log_{b}c}\right).\label{eq:cor2-expected-1}
\end{equation}
For each $m<m^{\star}$, $1-\alpha m\log_{b}c>0$, so the term
\begin{align*}
{n \choose m}\exp\left(-\frac{1}{2}n^{1-\alpha m\log_{b}c}\right) & \leq\exp\left(m\ln n-\frac{1}{2}n^{1-\alpha m\log_{b}c}\right)\\
 & =o(1)
\end{align*}
Hence, each of the $O\left(1\right)$ terms inside \prettyref{eq:cor2-expected-1}
is $o\left(1\right)$. It follows that $E[X]=o\left(1\right)$, so
by the first moment method, a.a.s. $X=0$. 

This proves a). 

To prove b), suppose $m>m^{\star}$, and recall $h^{\star}=\left(\frac{1}{2}-\epsilon\right)\frac{\ln\ln n}{\ln b}$.
Let $S=S\left(M\right)$ be the minimum complete set containing $M$.
To show that $\mathbf{E_{2}}$ holds for $M$, we need to show that
$\forall u\in S\left(M,h^{\star}\right)\setminus S(M)$, $e(u,M)\leq\alpha m$.
We wish to choose a new complete subset $S^{\prime}$ that also contains
$M$, and whose height $h^{\prime}$ is large enough that the vertices
in $S\left(M,h^{\star}\right)\setminus S^{\prime}$ are likely to
have at most $\alpha m$ edges to vertices in $M$. More explicitly,
for our choice of $h^{\prime}$ and $S^{\prime}$, let $\mathbf{A}$
be the event that $\forall u\in S^{\prime}\setminus S(M)$, $e(u,M)\leq\alpha m$,
and $\mathbf{B}$ be the event that $\forall u\in S\left(M,h^{\star}\right)\setminus S$,
$e(u,M)\leq\alpha m$. Then the events $\mathbf{A}$ and $\mathbf{B}$
are disjoint and independent, and $\mathbf{E_{2}}\left(M\right)=\mathbf{A}\cap\mathbf{B}$. 

In particular, we will choose $h^{\prime}$ to be larger than some
constant $\delta=\delta\left(\alpha,b,c\right)$. If $m$ (and hence
$h$) is increasing with $n$ then surely we have $h>\delta$. In
this case, taking $S^{\prime}=S(M)$ and $h^{\prime}=h$, we have
that $\mathbf{A}$ trivially occurs. If $m$ and $h$ are constant,
then $h^{\prime}$ will also be a constant that is possibly larger
than $h$. Since $S^{\prime}$ is of constant size, there are a constant
number of vertices in $S^{\prime}\setminus S(M)$. Furthermore, each
of these vertices will have at most $\alpha m$ neighbours in $M$
with some constant probability, because $M$ is of constant size.
Hence, $\mathbf{A}$ will occur with at least constant probability,
say $\Pr\left(\mathbf{A}\right)\geq\gamma$. 

Thus, it will suffice to show the event $\mathbf{B}$ occurs with
at least constant probability. Given a vertex $u$ of height $j>h^{\prime}$
from $M$, there is a uniform probability $c^{-j}$ of an edge between
$u$ and a particular vertex in $M$. Let $X_{u}=\mbox{Bin}\left(m,c^{-j}\right)$
be the number of such links. Since, $j>h^{\prime}>\delta$, by taking
$\delta=\delta\left(\alpha,b,c\right)$ to be large enough, we can
require that $c^{-j}\leq\frac{\alpha}{2}$. Hence, $\alpha m>2mc^{-j}$
and it follows from \prettyref{eq:L4-ptool-simple-intermediate} with
$s=\alpha m$ that

\[
\Pr\left(X_{u}\geq\alpha m\right)<2\left(\frac{e}{\alpha c^{j}}\right)^{\alpha m}.
\]

Now let $\mathbf{R_{j}}$ be the event that there exists a $u\in G$
of height $j>h^{\prime}$ from $M$ such that $X_{u}\geq\alpha m$.
There are fewer than $b^{j}$ such vertices, so by the union bound,
\begin{align*}
\Pr\left(\mathbf{R_{j}}\right) & \leq b^{j}2\left(\frac{e}{\alpha c^{j}}\right)^{\alpha m}.
\end{align*}

$\mathbf{B}$ occurs only if each of the disjoint, independent events
$\mathbf{R_{j}}$, $h^{\prime}<j\leq h^{\star}$, does \emph{not}
occur. Therefore

\begin{align}
\Pr\left(\mathbf{B}\right) & \geq1-\sum_{j>h^{\prime}}\Pr\left(\mathbf{R_{j}}\right)\nonumber \\
 & \geq1-2e^{\alpha m}\left(\frac{c^{\alpha m}}{c^{\alpha m}-b}\right)\left(\frac{b}{c^{\alpha m}}\right)^{h^{\prime}}\label{eq:E3lowerbd-1}
\end{align}

Now, it suffices to show that 

\[
2e^{\alpha m}\left(\frac{c^{\alpha m}}{c^{\alpha m}-b}\right)\left(\frac{b}{c^{\alpha m}}\right)^{h^{\prime}}<\frac{1}{2}
\]
which is true when 

\begin{equation}
h^{\prime}>\frac{4\alpha m\left(1+\ln c\right)-\ln\left(c^{\alpha m}-b\right)}{\alpha m\ln c-\ln b}\label{eq:h^prime bound-1-1}
\end{equation}
holds. We will see that this holds for $h^{\prime}>\delta$ when $\delta=\delta\left(\alpha,b,c\right)$
is chosen to sufficiently sufficiently large. Because $m$ is integer-valued
and larger than $m^{\star}$, $c^{\alpha m}-b$ is bounded below by
a positive number, and so for $m=\Theta(1)$, each term in both the
numerator and denominator of \prettyref{eq:h^prime bound-1-1} are
bounded. On the other hand, if $m$ is increasing, then we will have
for large enough $m$ that 
\[
\frac{4\alpha m\left(1+\ln c\right)-\ln\left(c^{\alpha m}-b\right)}{\alpha m\ln c-\ln b}\leq8\frac{1+\ln c}{\ln c}
\]
so that the right side of \prettyref{eq:h^prime bound-1-1} does not
depend on $m$.

Hence, \prettyref{eq:E3lowerbd-1} gives that $\Pr\left(\mathbf{B}\right)\geq\frac{1}{2}$,
which suffices to prove b).
\end{proof}
The next lemma proves \prettyref{eq:size_of_M}:

\newtheorem*{lem6}{Lemma}
\begin{lem6} \label{lem:E_3_occurs}Let $m^{\star}=\frac{\ln b}{\alpha\ln c}$ and $h^{\star}=\left(\frac{1}{2}-\epsilon\right)\frac{\ln\ln n}{\ln b}$. Let $\mathcal{M}$ be a family of sets such that each set $M\in\mathcal{M}$ is of size $m>m^{\star}$ and height $h\leq h^{\star}$. Furthermore, suppose for every $M_{1},M_{2}\in\mathcal{M}$, the complete sets of height $h^{\star}$, $S\left(M_{1},h^{\star}\right)$ and $S\left(M_{2},h^{\star}\right)$, do not intersect, and no edges between $S\left(M_{1},h^{\star}\right)$ and $S\left(M_{2},h^{\star}\right)$ have yet been exposed. (For each $M\in\mathcal{M}$, we allow any number of internal edges in $S\left(M,h^{\star}\right)$ to have been already exposed.)
Then a.a.s. at least  \[ \min\left(\left|\mathcal{M}\right|,\left(\ln n\right)^{\frac{\alpha\ln c}{4\ln b}\left(m-m^{\star}\right)}\right) \]
sets in $\mathcal{M}$ have $\mathbf{E_{3}}$.
\end{lem6}
\begin{proof}
Consider a set $M\in\mathcal{M}$ so that $m=\left|M\right|$. Given
a vertex $u\notin M$ of height $j>h^{\star}$ from $M$, there is
a probability $c^{-j}$ of an edge between $u$ and a particular vertex
in $M$. Let $X_{u}=\mbox{Bin}\left(m,c^{-j}\right)$ be the total
number of such edges. Since $c^{-j}=o(1)$, $\alpha m>2mc^{-j}$ and
it follows from \prettyref{eq:L4-ptool-simple-intermediate} with
$s=\alpha m$ that

\[
\Pr\left(X_{u}\geq\alpha m\right)<2\left(\frac{e}{\alpha c^{j}}\right)^{\alpha m}.
\]

Now let $\mathbf{A_{j}}$ be the event that there exists a $u\in G$
of height $j>h^{\star}$ from $M$ such that $X_{u}\geq\alpha m$.
There are $\left(b-1\right)b^{j-1}\leq b^{j}$ such vertices, so by
the union bound, 
\begin{align*}
\Pr\left(\mathbf{A_{j}}\right) & \leq b^{j}2\left(\frac{e}{\alpha c^{j}}\right)^{\alpha m}.
\end{align*}

$\mathbf{E_{3}}\left(M\right)$ occurs if and only if each of the
disjoint, independent events $\mathbf{A_{j}}$, $j>h^{\star}$, does
\emph{not} occur. Therefore, since $m>m^{\star}$ implies that $\frac{b}{c^{\alpha m}}<1$,
we have by the union bound that

\begin{align}
\Pr\left(\mathbf{E_{3}}\left(M\right)\right) & \geq1-\sum_{j>h^{\star}}\Pr\left(\mathbf{A_{j}}\right)\nonumber \\
 & =1-2e^{\alpha m}\sum_{j>h^{\star}}\left(\frac{b}{c^{\alpha m}}\right)^{j}\nonumber \\
 & \geq1-\frac{2e^{\alpha m}\left(\frac{b}{c^{\alpha m}}\right)^{h^{\star}}}{1-\left(\frac{b}{c^{\alpha m}}\right)}\label{eq:E_3 lower bound}
\end{align}

Now, considering the term \textrm{$e^{\alpha m}\left(\frac{b}{c^{\alpha m}}\right)^{h^{\star}}=b^{h^{\star}}\left(\frac{e}{c^{h^{\star}}}\right)^{\alpha m}$
}in the numerator of \prettyref{eq:E_3 lower bound}, we have:

\begin{align*}
b^{h^{\star}}\left(\frac{e}{c^{h^{\star}}}\right)^{\alpha m} & =\exp\left(\alpha m+h^{\star}\left(\ln b-\alpha m\ln c\right)\right)\\
 & =\exp\left(\alpha m-\left(\frac{1}{2}-\epsilon\right)\alpha\frac{\ln c}{\ln b}\left(m-m^{\star}\right)\ln\ln n\right)\\
 & \leq\exp\left(-\frac{\alpha\ln c}{2.5\ln b}\left(m-m^{\star}\right)\ln\ln n\right).
\end{align*}
Since $1-\left(\frac{b}{c^{\alpha m}}\right)$ is a constant, we have
from \prettyref{eq:E_3 lower bound} that 
\begin{align}
\Pr\left(\mathbf{E_{3}}\left(M\right)\right) & \geq1-\exp\left(-\frac{\alpha\ln c}{3\ln b}\left(m-m^{\star}\right)\ln\ln n\right).\label{eq:E_3_loewr_bound_final}
\end{align}

Because removing edges only increases the probability that $\mathbf{E_{3}}$
occurs, it is a monotone property. More precisely, let $G_{1}$ and
$G_{2}$ be graphs on the same vertex set $V$ and let $E\left(G_{1}\right)\subset E\left(G_{2}\right)$.
For any set $M\subset V$, if $\mathbf{E_{3}}\left(M\right)$ holds
in $G_{2}$ then it also holds in $G_{1}$. Hence, for two sets $M_{1}$
and $M_{2}$, Proposition 6.3.1 in \cite{Alon2000} implies that $\mathbf{E_{3}}\left(M_{1}\right)$
and $\mathbf{E_{3}}\left(M_{2}\right)$ are positively correlated;
that is:

\begin{equation}
\Pr\left(\mathbf{E_{3}}\left(M_{1}\right)\cap\mathbf{E_{3}}\left(M_{2}\right)\right)\geq\Pr\left(\mathbf{E_{3}}\left(M_{1}\right)\right)\Pr\left(\mathbf{E_{3}}\left(M_{2}\right)\right).\label{eq:pos-correlation}
\end{equation}

Now, let $M_{1},M_{2},\dots$ be any ordering of the sets in $\mathcal{M}$.
We will expose the edges necessary for $\mathbf{E_{3}}\left(M_{i}\right)$
using this order. \prettyref{eq:pos-correlation} imples that for
any $i$, 
\[
\Pr\left(\mathbf{E_{3}}\left(M_{i}\right)|\cap_{j<i}\mathbf{E_{3}}\left(M_{j}\right)\right)\geq\Pr\left(\mathbf{E_{3}}\left(M_{i}\right)\right).
\]

The union bound and \prettyref{eq:E_3_loewr_bound_final} thus implies
that for some value $k$, 
\begin{align*}
\Pr\left(\cap_{i<k}\mathbf{E_{3}}\left(M_{i}\right)\right) & \geq1-\sum_{i<k}\Pr\left(\lnot\mathbf{E_{3}}\left(M_{i}\right)\right)\\
 & \geq1-k\exp\left(-\frac{\alpha\ln c}{3\ln b}\left(m-m^{\star}\right)\ln\ln n\right)
\end{align*}

Hence, if 
\[
\ln k<\frac{\alpha\ln c}{3\ln b}\left(m-m^{\star}\right)\ln\ln n
\]
then it follows that a.a.s. every set $M_{i}$ for $1\leq i\leq k$
has the property $\mathbf{E_{3}}\left(M_{i}\right)$. Taking $k=\min\left(\left|\mathcal{M}\right|,\left(\ln n\right)^{\frac{\alpha\ln c}{4\ln b}\left(m-m^{\star}\right)}\right)$
proves the lemma.
\end{proof}
In the proof of \prettyref{cor:Constant-size-existence-1}, the expected
number of sets with $\mathbf{D}\cap\mathbf{E_{1}}\cap\mathbf{E_{2}}$
is $qn/b^{h^{\star}}=\Omega\left(n^{3/4}\right)$. Setting $t=n^{2/3}$
in \prettyref{lem:Bollobas-prob-tool} gives that a.a.s. there is
a subset $\mathcal{M}^{\prime}\subset\mathcal{M}$ of size at least
$\Omega\left(n^{3/4}\right)-n^{2/3}=\Omega\left(n^{3/4}\right)$ such
that each set in $\mathcal{M}^{\prime}$ has $\mathbf{D}\cap\mathbf{E_{1}}\cap\mathbf{E_{2}}$. 

Next, we give the proof of \prettyref{thm:Thm1}a) in more detail,
with the omitted calculations inserted:

\newtheorem*{thm1a}{Theorem 1a}
\begin{thm1a} Let $m^{\star}=\frac{\ln b}{\alpha\ln c}$. There a.a.s. exists clusters of size larger than $\left(\ln n\right)^{\frac{1}{2}-\epsilon}$ in $G$. Moreover, there exists a constant $\gamma=\gamma\left(\alpha,b,c\right)>0$ such that for each $h$ satisfying $\log_{b}m^{\star}<h\leq\left(\frac{1}{2}-\epsilon\right)\frac{\ln\ln n}{\ln b}$, there a.a.s. exists at least $\left(\ln n\right)^{\gamma b^{h}}$ complete clusters of size $b^{h}$.\end{thm1a}
\begin{proof}
Note that the second part of the theorem statement implies the first,
because a complete set of height $h^{\star}=\left(\frac{1}{2}-\epsilon\right)\frac{\ln\ln n}{\ln b}$
has $b^{h^{\star}}=\left(\ln n\right)^{\frac{1}{2}-\epsilon}$ vertices.
Let $M$ be a complete set of height $h$. We will first examine the
event $\mathbf{D}$ that $M$ is internally dense. 

In fact, let us consider the stronger event of $M$ being a clique.
Each potential edge in $M$ occurs with probability at least $c^{-h}$.
so the probability of all ${b^{h} \choose 2}$ edges occuring is at
least

\[
\left(c^{-h}\right)^{{b^{h} \choose 2}}\geq c^{-hb^{2h}}
\]
Hence, $\Pr\left(\mathbf{D}\right)\geq\exp\left(-\ln\left(c\right)hb^{2h}\right)$. 

Now, partition $G$ into complete sets of height $h^{\star}=\left(\frac{1}{2}-\epsilon\right)\frac{\ln\ln n}{\ln b}$.
From each such set, choose a single complete set of height $h$, and
let these $n/b^{h^{\star}}$ disjoint sets form the family $\mathcal{M}$.
Because these sets are complete, they automatically have $\mathbf{E_{1}}$.
Because they do not overlap and lie in different sets of height $h^{\star}$,
they will have $\mathbf{D}$ and $\mathbf{E_{2}}$ independently.
\prettyref{lem:E_1_and__E2-1} b) implies they have $\mathbf{E_{2}}$
with probability at least $a$ for some positive constant $a$. Hence,
for every $M\in\mathcal{M}$, we have that
\begin{align}
\Pr\left(\mathbf{D}\cap\mathbf{E_{1}}\cap\mathbf{E_{2}}\right) & \geq\Pr\left(\mathbf{D}\right)\Pr\left(\mathbf{E_{1}}\right)\Pr\left(\mathbf{E_{2}}\right)\nonumber \\
 & \geq a\exp\left(-\ln\left(c\right)hb^{2h}\right)\label{eq:D_E1_E2_lower_bd-2}
\end{align}
Now, the number of sets in $\mathcal{M}$ with $\mathbf{D}\cap\mathbf{E_{1}}\cap\mathbf{E_{2}}$
stochastically dominates the random variable $X=\mbox{Bin}\left(\frac{n}{b^{h^{\star}}},a\exp\left(-\ln\left(c\right)hb^{2h}\right)\right)$.
Since $h\leq h^{\star}$, we have that 
\begin{align*}
E[X] & \geq\frac{n}{b^{h^{\star}}}a\exp\left(-\ln\left(c\right)hb^{2h}\right)\\
 & =n\exp\left(\ln a-\ln\left(b\right)h^{\star}-\ln\left(c\right)hb^{2h}\right)\\
 & \geq n\exp\left(-h^{\star2}b^{2h^{\star}}\right)
\end{align*}

Since $h^{\star}=\left(\frac{1}{2}-\epsilon\right)\frac{\ln\ln n}{\ln b}$,
this becomes 

\begin{align*}
E[X] & \geq\frac{n}{\left(\frac{\frac{1}{2}-\epsilon}{\ln b}\right)^{2}\left(\ln\ln n\right)^{2}\left(\ln n\right)^{1-2\epsilon}}
\end{align*}

Notice that $E[X]$ is asymptotically larger than $\frac{n}{\ln n}$.
Setting $t=n^{2/3}$ so that $E[X]-t\geq\frac{n}{\ln n}$, \prettyref{lem:Bollobas-prob-tool}
gives that 
\begin{align*}
\Pr\left(X<\frac{n}{\ln n}\right) & \leq\exp\left(-\frac{n^{4/3}}{2E[X]}\right)\\
 & \leq\exp\left(n^{-1/3}\right)
\end{align*}

In other words, a.a.s. there are at least $\frac{n}{\ln n}$ sets
in $\mathcal{M}$ with $\mathbf{D}\cap\mathbf{E_{1}}\cap\mathbf{E_{2}}$.
Let $\mathcal{M}^{\prime}\subset\mathcal{M}$ be this sub-family of
sets. Because each set in $\mathcal{M}^{\prime}$ lies in a different
complete set of height $h^{\star}$, and we have only exposed edges
inside these complete sets of height $h^{\star}$, the conditions
of Lemma \ref{lem:lemm_6_E3} apply to $\mathcal{M}^{\prime}$. Hence
at least $\left(\ln n\right)^{\frac{\alpha\ln c}{4\ln b}\left(b^{h}-m^{\star}\right)}$
sets in $\mathcal{M}^{\prime}$ also have $\mathbf{E_{3}}$, and therefore
are clusters.

It remains to find a constant $\gamma=\gamma\left(\alpha,b,c\right)>0$
such that there are a.a.s. at least $\left(\ln n\right)^{\gamma b^{h}}$clusters
of size $b^{h}$. Recall that $m^{\star}=\frac{\ln b}{\alpha\ln c}$,
and let $h_{min}$ be the minimum integral value of $h$ such that
$b^{h}>m^{\star}$; setting 
\[
\gamma=\frac{\alpha\ln c}{4\ln b}\frac{\left(b^{h_{min}}-m^{\star}\right)}{b^{h_{min}}}
\]
guarantees that $\left(\ln n\right)^{\gamma b^{h}}\leq\left(\ln n\right)^{\frac{\alpha\ln c}{4\ln b}\left(b^{h}-m^{\star}\right)}$,
and hence there are a.a.s. at least $\left(\ln n\right)^{\gamma b^{h}}$
clusters of height $h$ in $G$.
\end{proof}
\newtheorem*{lem8}{Lemma 5}
\begin{lem8} For all $\epsilon$ such that $\frac{\ln c}{4\ln b}>\epsilon>0$, a.a.s. for each short $\epsilon$-thick set $M$, there exists a set $M_{1}\subset M$ such that $\left|M_{1}\right|\geq\frac{3}{4} \left|M\right|$ and $\forall v\in M_{1}$, $e(v,M)\leq\frac{\beta}{4} \left| M \right|$.\end{lem8}
\begin{proof}
Let $m=\left|M\right|$, and let $S=S\left(M,h_{\epsilon}\right)$
be the complete set of height $h_{\epsilon}=\left(\frac{1}{2}+\epsilon\right)\frac{\ln\ln n}{\ln b}$
containing $M$. If $S$ contains less than $\frac{\beta}{32}m^{2}$
internal edges, then it follows that there can be no set $M^{\prime}\subset M$
of size $\left|M^{\prime}\right|\geq\frac{m}{4}$ such that $\forall v\in M^{\prime}$,
$e(v,M)\geq\frac{\beta}{4}m$. Hence, it suffices to show that a.a.s.
all complete sets $S$ of height $h_{\epsilon}$ have less than $\frac{\beta}{32}m^{2}$
internal edges.

Consider a complete set of height $j$. It contains $b$ sets of height
$j-1$, each of which contains $b^{j-1}$ vertices. Hence, there are
${b \choose 2}b^{2\left(j-1\right)}$ potential edges of height $j$
in a single complete set of height $j$. Let the actual number of
such edges be given by $X_{j}$; since each occurs with probability
$c^{-j}$, we have that 
\[
E[X_{j}]={b \choose 2}b^{2\left(j-1\right)}c^{-j}.
\]

Now, let $X_{S}$ be the number of edges in $S$. We wish to show
that $E[X_{S}]=O\left(\ln n\right)$. Since there are $b^{h_{\epsilon}-j}$
complete sets of height $j$ in $S$, we have that 
\begin{align}
E[X_{S}] & =\sum_{j=1}^{h_{\epsilon}}b^{h_{\epsilon}-j}E[X_{j}]\nonumber \\
 & =\frac{b-1}{2b}b^{h_{\epsilon}}\sum_{j=1}^{h_{\epsilon}}\left(\frac{b}{c}\right)^{j}\label{eq:lem8_EX_S_sum}
\end{align}

The sum in this expression is bounded as follows:

\begin{align*}
\sum_{j=1}^{h_{\epsilon}}\left(\frac{b}{c}\right)^{j}\leq & \begin{cases}
\frac{c}{c-b} & \mbox{if }b<c\\
h_{\epsilon} & \mbox{if }b=c\\
\frac{b}{b-c}\left(\frac{b}{c}\right)^{h_{\epsilon}+1} & \mbox{if }b>c
\end{cases}
\end{align*}

In the first two cases, this sum is small enough that \prettyref{eq:lem8_EX_S_sum}
combined with the fact that $b^{h_{\epsilon}}=\left(\ln n\right)^{\frac{1}{2}+\epsilon}$
easily gives that $E[X_{S}]=O\left(\ln n\right)$. In the case that
$b>c$, we may rewrite \prettyref{eq:lem8_EX_S_sum} as follows:
\begin{align}
E[X_{S}]\leq & \frac{b\left(b-1\right)}{2c\left(b-c\right)}b^{h_{\epsilon}}\left(\frac{b}{c}\right)^{h_{\epsilon}}\nonumber \\
= & O\left(b^{h_{\epsilon}}\left(\frac{b}{c}\right)^{h_{\epsilon}}\right)\label{eq:EX_S bound b geq c}
\end{align}

Now, since $h_{\epsilon}=\left(\frac{1}{2}+\epsilon\right)\frac{\ln\ln n}{\ln b}$
, we have $b^{h_{\epsilon}}=\left(\ln n\right)^{\frac{1}{2}+\epsilon}$
and $\left(\frac{b}{c}\right)^{h_{\epsilon}}=\left(\ln n\right)^{\left(\frac{1}{2}+\epsilon\right)\left(1-\frac{\ln c}{\ln b}\right)}$.
Since $\epsilon<\frac{\ln c}{4\ln b}$, it follows that $\left(1+2\epsilon\right)\left(1-\frac{\ln c}{\ln b}\right)<1$,
so 

\begin{align*}
b^{h_{\epsilon}}\left(\frac{b}{c}\right)^{h_{\epsilon}} & =\left(\ln n\right)^{\left(1+2\epsilon\right)\left(1-\frac{\ln c}{\ln b}\right)}\\
 & \leq\ln n.
\end{align*}

Hence, from \prettyref{eq:EX_S bound b geq c} we have $E[X_{S}]=O\left(\ln n\right)$
holds as well when $b>c$, and therefore holds in all cases.

We now use this bound on the expected value of $X_{S}$ to bound the
probability that $X_{S}\geq\frac{\beta}{32}m^{2}=\frac{\beta}{32}\left(\ln n\right)^{1+\frac{2\epsilon}{3}}$.
Set $t=\left(\ln n\right)^{\left(1+\frac{\epsilon}{4}\right)}$, so
that $E[X_{S}]+t<2\left(\ln n\right)^{1+\frac{\epsilon}{4}}\leq\frac{\beta}{32}m^{2}$,
since $m\geq\left(\ln n\right)^{\frac{1}{2}+\frac{\epsilon}{3}}$.
Applying \prettyref{lem:Bollobas-prob-tool} with this $t$, we have
\[
\Pr\left(X_{S}\geq\frac{\beta}{32}m^{2}\right)\leq\exp\left(-\frac{1}{4}\left(\ln n\right)^{1+\frac{\epsilon}{4}}\right).
\]

There are $\frac{n}{b^{h_{\epsilon}}}\leq n$ complete sets of height
$h_{\epsilon}$. Letting $X$ denote the number of such sets with
at least $\frac{\beta}{32}m^{2}$ internal edges, it follows that
\begin{align*}
E[X] & =n\Pr\left(X_{S}\geq\frac{\beta}{32}m^{2}\right)\\
 & \leq\exp\left(\ln n-\frac{1}{4}\left(\ln n\right)^{1+\frac{\epsilon}{4}}\right).
\end{align*}

This goes to zero as $n\rightarrow\infty$, showing that a.a.s. $X=0$.
\end{proof}
\newtheorem*{lem9}{Lemma 6}
\begin{lem9} For all $\epsilon>0$, a.a.s. for every tall $\epsilon$-thick set $T=M\cup K$, where $M$ is a short $\epsilon$-thick and $K$ is a set with $\left|K\right|\geq\frac{\beta}{4}\left|M\right|$ such that $K$ does not intersect $S\left(M,h_{\epsilon}\right)$, the complete set of height $h_{\epsilon}=\left(\frac{1}{2}+\epsilon\right)\frac{\ln\ln n}{\ln b}$, there exists a set $M_{2}\subset M$ such that $\left|M_{2}\right|>\frac{3}{4}\left|M\right|$ and $\forall v\in M$, $e\left(v,K\right)\leq\frac{\beta}{2}\left|K\right|$. \end{lem9}
\begin{proof}
Let $m=\left|M\right|$ and $k=\left|K\right|$, and let $M$ and
$K$ be such that the conditions of the lemma hold. Since vertices
in $K$ are at height at least $h_{\epsilon}$ from $M$, each potential
edge between any $v\in M$ and $K$ occurs independently with probability
at most $c^{-h_{\epsilon}}$. Hence, $E\left[e\left(v,K\right)\right]\leq kc^{-h_{\epsilon}}<\frac{\beta k}{4}$
for large enough $n$. Setting $t=\frac{\beta k}{4}$, by \prettyref{lem:Bollobas-prob-tool}
we have that

\begin{align}
\Pr\left(e\left(v,K\right)>\frac{\beta k}{2}\right) & \leq\exp\left(-\frac{3}{8}t\right).\nonumber \\
 & \leq\exp\left(-\frac{\beta}{16}k\right)\label{eq:L7_Pr_bd_on_Xv}
\end{align}

Now, let $M_{2}=\left\{ v\in M:\: e\left(v,K\right)\leq\frac{\beta k}{2}\right\} $,
and let $X_{M,K}=\left|M-M_{2}\right|$. We wish to show that a.a.s.
for all valid choices of $M$ and $K$, $X_{M,K}\leq\frac{m}{4}$.
Let $p_{M,K}$ denote the probability that $X_{M,K}>\frac{m}{4}$.
From \prettyref{eq:L7_Pr_bd_on_Xv}, it follows that $X_{M,K}$ is
stochastically dominated by the random variable $\mbox{Bin}\left(m,\exp\left(-\frac{\beta}{16}k\right)\right)$.
Therefore, by \prettyref{eq:L4-prob-tool-simple}, setting $s=\frac{m}{4}$,
we have that

\begin{align*}
p_{M,K} & \leq2\exp\left(\frac{m}{4}\left(\ln m+1-\ln\left(\frac{m}{4}\right)-\frac{\beta}{16}k\right)\right)\\
 & \leq\exp\left(-\frac{\beta}{100}mk\right).
\end{align*}

Now, let $S$ be a complete set of height $\frac{\left(\ln n\right)^{\frac{1}{2}}}{\ln b}$;
we will count the expected number $X_{S}$ of tall $\epsilon$-thick
sets $T=M\cup K$ inside $S$ such that $X_{M,K}>\frac{m}{4}$. We
have that $E[X_{S}]\leq\sum_{M,K}p_{M,K}$, where this sum ranges
over all valid choices of $T=M\cup K$ inside $S$. Fixing the set
$M$ and a size $k$, we have that there are at most 
\[
{\left|S\right| \choose k}={b^{\frac{(\ln n)^{\frac{1}{2}}}{\ln b}} \choose k}\leq\exp\left(k\left(\ln n\right)^{\frac{1}{2}}\right)
\]
sets $K$ of size $k$; hence, 
\begin{align*}
E[X_{S}] & \leq\sum_{M}\sum_{k}\exp\left(k\left(\left(\ln n\right)^{\frac{1}{2}}-\frac{\beta}{100}m\right)\right).
\end{align*}

Since $m\geq\left(\ln n\right)^{\frac{1}{2}+\frac{\epsilon}{3}}$,
the term in the exponential is negative, and hence is maximized when
$k$ is minimized; that is, when $k=\frac{\beta}{4}m$. Furthermore,
the height of $K$ is at most $\frac{\left(\ln n\right)^{\frac{1}{2}}}{\ln b}$,
so $k\leq\exp\left(\left(\ln n\right)^{\frac{1}{2}}\right)$. Hence,
\begin{align*}
E[X_{S}] & \leq\sum_{M}\sum_{k}\exp\left(\frac{\beta m}{4}\left(\left(\ln n\right)^{\frac{1}{2}}-\frac{\beta}{100}m\right)\right)\\
 & \sum_{M}\exp\left(\left(\ln n\right)^{\frac{1}{2}}\right)\exp\left(\frac{\beta m}{4}\left(\left(\ln n\right)^{\frac{1}{2}}-\frac{\beta}{100}m\right)\right)\\
 & \sum_{M}\exp\left(-\frac{\beta^{2}}{500}m^{2}\right)
\end{align*}
since $\left(\ln n\right)^{\frac{1}{2}}=o\left(m\right)$.

Now, within $S$ there are $b^{\frac{\left(\ln n\right)^{\frac{1}{2}}}{\ln b}-h_{\epsilon}}\leq\exp\left(\left(\ln n\right)^{\frac{1}{2}}\right)$
complete sets of height $h_{\epsilon}$, and each has at most ${\left(\ln n\right)^{\frac{1}{2}+\epsilon} \choose m}\leq\exp\left(m\ln\ln n\right)$
subsets $M$ of size $m$, so there are at most $\exp\left(\left(\ln n\right)^{\frac{1}{2}}+m\ln\ln n\right)\leq\exp\left(2m\ln\ln n\right)$
choices total for the set $M$ within $S$. Hence,

\begin{align*}
E[X_{S}] & \leq\sum_{m}\exp\left(2m\ln\ln n\right)\exp\left(-\frac{\beta^{2}m^{2}}{500}\right)\\
 & \leq\sum_{m}\exp\left(-\frac{\beta^{2}m^{2}}{1000}\right)
\end{align*}

since $m\ln\ln n=o\left(m^{2}\right)$. Again, the term in the sum
is maximized when $m$ is minimized; since $\left(\ln n\right)^{\frac{1}{2}+\frac{\epsilon}{3}}\leq m\leq\left(\ln n\right)^{\frac{1}{2}+\epsilon}$,
it follows

\begin{align}
E[X_{S}] & \leq\left(\ln n\right)^{\frac{1}{2}+\epsilon}\exp\left(-\frac{\beta^{2}}{1000}\left(\ln n\right)^{1+\frac{2\epsilon}{3}}\right)\nonumber \\
 & \leq\exp\left(-\frac{\beta^{2}}{2000}\left(\ln n\right)^{1+\frac{2\epsilon}{3}}\right).\label{eq:l10finalstep}
\end{align}

Now, there are fewer than $n$ choices for the complete set $S$.
Let $X=\sum_{S}X_{S}$ denote the total number of tall $\epsilon$-thick
sets $T=M\cup K$ such that $X_{M,K}>\frac{m}{4}$. Then $E[X]\leq nE[X_{S}]$.
Since $\ln n=o\left(\left(\ln n\right)^{1+\frac{2\epsilon}{3}}\right)$,
\prettyref{eq:l10finalstep} implies that $E[X]=o(1)$, and hence
a.a.s. $X=0$.
\end{proof}
\newtheorem*{lem11}{Lemma 7}
\begin{lem11} For all $\epsilon$ such that $\frac{\ln c}{4\ln b}>\epsilon>0$, a.a.s. for every tall $\epsilon$-thick set $T$, there exists a set $T^{\prime}\subset T$ such that $\left|T^{\prime}\right|>\left(\ln n\right)^{\frac{1}{2}+\frac{\epsilon}{4}}$ and $\forall v\in T^{\prime}$, $e\left(v,T\right)\leq\frac{\beta}{2}\left|T\right|$. \end{lem11}
\begin{proof}
We consider two cases: either $T$ contains a short $\epsilon$-thick
set as a subset, or it does not. Let $t=\left|T\right|$.

First, suppose $T$ contains a short $\epsilon$-thick set $M$. Adding
vertices in $S\left(M,h_{\epsilon}\right)\setminus M$ to $M$ will
preserve the property that $M$ is a short $\epsilon$-thick set,
so we may assume that $M$ is maximal. In other words, we may assume
the set $T\setminus M$ does not intersect the compelte set of height
$h_{\epsilon}$, $S\left(M,h_{\epsilon}\right)$. Set $K=T\setminus M$,
and Let $m=\left|M\right|$ and $k=\left|K\right|$. By \prettyref{lem:thick-cluster-upper-bd},
there is a set $M_{1}\subset M$ such that $\left|M_{1}\right|\geq\frac{3}{4}m$
and $\forall v\in M_{1}$, $e\left(v,M\right)\leq\frac{\beta}{4}m$.
If $k<\frac{\beta m}{4}$, then even if every edge from $v$ to $K$
exists, we will still have $e\left(v,M\cup K\right)\leq\frac{\beta}{4}m+k\leq\frac{\beta}{2}\left(m+k\right)$
for every $v\in M_{1}$, and hence may choose $T^{\prime}=M_{1}$.
If $k\geq\frac{\beta m}{4}$, then by \prettyref{lem:L7-NoC_K sets with many bad vertices},
there is some set $M_{2}\subset M$ such that $\left|M_{2}\right|>\frac{3}{4}m$
and $\forall v\in M_{2}$, $e\left(v,K\right)\leq\frac{\beta}{2}k$.
Taking $T^{\prime}=M_{1}\cap M_{2}$, it follows that $\left|T^{\prime}\right|\geq\frac{m}{2}$
and $e\left(v,M\cup K\right)\leq\frac{\beta}{2}m+\frac{\beta}{2}k=\frac{\beta}{2}t$
for every $v\in T^{\prime}$. 

Otherwise, suppose $T$ contains no short $\epsilon$-thick sets.
That is, each set of height $h_{\epsilon}=\left(\frac{1}{2}+\epsilon\right)\frac{\ln\ln n}{\ln b}$
can contain at most $\left(\ln n\right)^{\frac{1}{2}+\frac{\epsilon}{3}}\leq\frac{\beta}{4}t$
vertices. For any $v\in T$, there are at least $\frac{3\beta}{4}t$
vertices in $T$ with height more than $h_{\epsilon}$ from $v$.
Let $X_{v}$ denote the number of edges from $v$ to $T$ of height
more than $h_{\epsilon}$. If $X_{v}\leq\frac{\beta}{4}t$, then certainly
$e(v,T)\leq\frac{\beta}{2}t$. Now, set $X_{T}=\frac{1}{2}\sum_{v\in T}X_{v}$
to be the total number of edges of height more than $h_{\epsilon}$
in $T$. If $X_{T}<\frac{\beta}{16}t^{2}$, then by a counting argument
there must be a set $T^{\prime}\subset T$ of size at least $\frac{t}{2}$
such that $\forall v\in T^{\prime}$, $X_{v}\leq\frac{\beta}{4}t$.
Hence, it suffices to show that a.a.s. for all $T$, $X_{T}<\frac{\beta}{16}t^{2}$.

Since there are less than ${t \choose 2}$ edges of height greater
than $h_{\epsilon}$ in $T$, each occurring with probability at most
$c^{-h_{\epsilon}}$, $X_{T}$ is stochastically dominated by the
random variable $\mbox{Bin}\left({t \choose 2},c^{-h_{\epsilon}}\right)$.
Since ${t \choose 2}c^{-h_{\epsilon}}=o\left(t^{2}\right),$we may
apply \prettyref{eq:L4-prob-tool-simple} with $s=\frac{\beta}{16}t^{2}$,
giving
\begin{align*}
\Pr\left(X_{T}\geq\frac{\beta}{16}t^{2}\right) & \leq2\exp\left(\frac{\beta}{16}t^{2}\left(\ln{t \choose 2}+1-\ln\left(\frac{\beta}{16}t^{2}\right)-h_{\epsilon}\ln c\right)\right)\\
 & \leq\exp\left(-\frac{\beta\ln c}{32}h_{\epsilon}t^{2}\right).
\end{align*}

Finally, let $X$ denote the number of sets $T$ such that $X_{T}\geq\frac{\beta}{16}t^{2}$.
Then since there are at most ${\exp\left(\left(\ln n\right)^{\frac{1}{2}}\right) \choose t}\leq\exp\left(t\left(\ln n\right)^{\frac{1}{2}}\right)$
sets $T$ of size $t$ contained within a single complete set of height
$\frac{\left(\ln n\right)^{\frac{1}{2}}}{\ln b}$, and there are less
than $n$ such complete sets, we have that 
\begin{align*}
E[X] & \leq n\sum_{t=\left(\ln n\right)^{\frac{1}{2}+\frac{\epsilon}{2}}}^{n}\exp\left(t\left(\left(\ln n\right)^{\frac{1}{2}}-\frac{\beta\ln c}{32}h_{\epsilon}t\right)\right)
\end{align*}

Since $t\geq\left(\ln n\right)^{\frac{1}{2}+\frac{\epsilon}{2}}$,
we have that $\left(\ln n\right)^{\frac{1}{2}}=o(th_{\epsilon})$.
Thus, the term in the exponential is negative and decreasing in $t$,
so it is maximized when $t=\left(\ln n\right)^{\frac{1}{2}+\frac{\epsilon}{2}}$.
The above becomes

\begin{align*}
E[X] & \leq n\sum_{t=\left(\ln n\right)^{\frac{1}{2}+\frac{\epsilon}{2}}}^{n}\exp\left(\left(\ln n\right)^{\frac{1}{2}+\frac{\epsilon}{2}}\left(\left(\ln n\right)^{\frac{1}{2}}-\frac{\beta\ln c}{32}\left(\ln n\right)^{\frac{1}{2}+\frac{\epsilon}{2}}h_{\epsilon}\right)\right)\\
 & \leq\exp\left(2\ln n-\left(\ln n\right)^{1+\epsilon}\right).
\end{align*}

Hence $E[X]=o(1)$, showing that a.a.s. $X_{T}<\frac{\beta}{16}t^{2}$
for every set $T$.
\end{proof}
\newtheorem*{lem12}{Lemma 8}
\begin{lem12} For all $\epsilon$ such that $\frac{\ln c}{4\ln b}>\epsilon>0$, there are a.a.s. no clusters $Q=T\cup R$ of size at least $\left(\ln n\right)^{\frac{1}{2}+\epsilon}$, where $T$ is a tall $\epsilon$-thick set. \end{lem12}
\begin{proof}
Let $t=\left|T\right|$ and $r=\left|R\right|$. $T$ is contained
within the complete set $S\left(T,\frac{\left(\ln n\right)^{\frac{1}{2}}}{\ln b}\right)$,
and adding vertices from $S\left(T,\frac{\left(\ln n\right)^{\frac{1}{2}}}{\ln b}\right)\setminus T$
to $T$ will preserve the property that $T$ is a tall $\epsilon$-thick
set. Hence, we may assume $T$ is maximal; that is, that the vertices
in $R$ are at height at least $\frac{\left(\ln n\right)^{\frac{1}{2}}}{\ln b}$
from the vertices in $T$. By \prettyref{lem:Tall_clusters_have_many_bad_vertices},
there exists a set $T^{\prime}\subset T$, $\left|T^{\prime}\right|\geq\left(\ln n\right)^{\frac{1}{2}+\frac{\epsilon}{4}}$,
such that $e\left(v,T\right)\leq\frac{\beta}{2}t$ for every $v\in T^{\prime}$.
For $T\cup R$ to be a cluster, we must have that $e\left(v,T\cup R\right)\geq\beta\left(t+r\right)$.
Hence, for $v\in T^{\prime}$, it follows that $e\left(v,R\right)\geq\frac{\beta}{2}t+\beta r$.
This implies first that $r\geq\frac{\beta}{2}t$; and second, that
we must have $e\left(v,R\right)\geq\beta r$ for every $v\in T^{\prime}$.
The remainder of the proof will show that a.a.s. no sets $T\cup R$
satisfy this condition.

Since $v\in T^{\prime}$ is height at least $\frac{\left(\ln n\right)^{\frac{1}{2}}}{\ln b}$
from each vertex in $R$, $e\left(v,R\right)$ is stochastically dominated
by the random variable $\mbox{Bin}\left(r,\exp\left(-\frac{\ln c}{\ln b}\left(\ln n\right)^{\frac{1}{2}}\right)\right)$.
By \prettyref{eq:L4-prob-tool-simple} with $s=\beta r$, we have

\begin{align*}
\Pr\left(e\left(v,R\right)\geq\beta r\right) & \leq2\exp\left(\beta r\left(1-\ln\beta-\frac{\ln c}{\ln b}\left(\ln n\right)^{\frac{1}{2}}\right)\right)\\
 & \leq\exp\left(-\frac{\beta\ln c}{2\ln b}r\left(\ln n\right)^{\frac{1}{2}}\right).
\end{align*}

Since the edges of $T^{\prime}\subset T$ are independent of the edges
between $T$ and $R$, this is true independently for all vertices
$v\in T^{\prime}$. Since $\left|T^{\prime}\right|\geq\left(\ln n\right)^{\frac{1}{2}+\frac{\epsilon}{4}}$,
the probability $p_{T}$ that $e\left(v,R\right)\geq\beta r$ holds
for all $v\in T^{\prime}$ is therefore bounded:

\begin{align*}
p_{T} & \leq\Pr\left(e\left(v,R\right)\geq\beta r\right)^{\left|T^{\prime}\right|}\\
 & \leq\exp\left(-\frac{\beta\ln c}{2\ln b}r\left(\ln n\right)^{1+\frac{\epsilon}{4}}\right).
\end{align*}

Now, let $X$ denote the total number of clusters of the form $T\cup R$.
There are at most ${n \choose r}\leq\exp\left(r\ln n\right)$ sets
of size $r$, so

\begin{align*}
E[X] & \leq\sum_{T}\sum_{r=\frac{\beta t}{2}}^{n}{n \choose r}p_{T}\\
 & \leq\sum_{T}\sum_{r}\exp\left(r\left(\ln n-\frac{\beta\ln c}{2\ln b}\left(\ln n\right)^{1+\frac{\epsilon}{4}}\right)\right).
\end{align*}

The term in the exponential is negative, and therefore maximized when
$r=\frac{\beta t}{2}$. For a fixed size $t$, there are less than
$n\,{e^{(\ln n)^{\frac{1}{2}}} \choose t}\leq n\,\exp\left(t\left(\ln n\right)^{\frac{1}{2}}\right)$
choices for the tall $\epsilon$-thick set $T$, and hence
\begin{align*}
E[X] & \leq\sum_{T}n\,\exp\left(\frac{\beta t}{2}\left(\ln n-\frac{\beta\ln c}{2\ln b}\left(\ln n\right)^{1+\frac{\epsilon}{4}}\right)\right)\\
 & \leq\sum_{T}\exp\left(-\frac{\beta^{2}\ln c}{8\ln b}t\left(\ln n\right)^{1+\frac{\epsilon}{4}}\right)\\
 & \leq\sum_{t=\left(\ln n\right)^{\frac{1}{2}}}^{n}\exp\left(t\left(\ln n\right)^{\frac{1}{2}}-\frac{\beta^{2}\ln c}{8\ln b}t\left(\ln n\right)^{1+\frac{\epsilon}{4}}\right).
\end{align*}

Since $t\left(\ln n\right)^{\frac{1}{2}}=o\left(t\left(\ln n\right)^{1+\frac{\epsilon}{4}}\right)$,
this exponential term is decreasing and hence maximized for the minimum
value of $t$; that is, for $t=\left(\ln n\right)^{\frac{1}{2}}$.
Thus

\begin{align*}
E[X] & \leq\sum_{t=\left(\ln n\right)^{\frac{1}{2}}}^{n}\exp\left(\ln n-\frac{\beta^{2}\ln c}{8\ln b}\left(\ln n\right)^{\frac{3}{2}+\frac{\epsilon}{4}}\right)\\
 & \leq n\,\exp\left(-\frac{\beta^{2}\ln c}{10\ln b}\left(\ln n\right)^{\frac{3}{2}+\frac{\epsilon}{4}}\right).
\end{align*}

Since $\ln n=o\left(\left(\ln n\right)^{\frac{3}{2}+\frac{\epsilon}{4}}\right)$
we have that $E[X]=o\left(1\right),$ implying that a.a.s. $X=0$. 
\end{proof}
Finally, we fill a detail from the proof of \prettyref{thm:Thm1}b).
To get \prettyref{eq:thm1b-bound-on-X_Q}, we use \prettyref{eq:L4-prob-tool-simple}
with $s=\frac{\beta}{4}q^{2}$, giving:

\begin{align*}
\Pr\left(X_{Q}\geq\frac{\beta}{4}q^{2}\right) & \leq2\exp\left(\frac{\beta}{4}q^{2}\left(\ln{q \choose 2}+1-\ln\left(\frac{\beta}{4}q^{2}\right)-h\ln c\right)\right)\\
 & \leq\exp\left(-\frac{\beta\ln c}{8}q^{2}h\right).
\end{align*}

\bibliographystyle{plain}
\bibliography{biblio}

\end{document}